\documentclass[apjl]{emulateapj}
\shorttitle{Tidal  break-up  of binary stars by a SMBH}
\shortauthors{Antonini et al.}

\begin{document}

\def\gap{\;\rlap{\lower 2.5pt
    \hbox{$\sim$}}\raise 1.5pt\hbox{$>$}\;}
\def\lap{\;\rlap{\lower 2.5pt
   \hbox{$\sim$}}\raise 1.5pt\hbox{$<$}\;}

\newcommand\msun{\rm\,M_\odot}
\newcommand\kms{{\rm\,km\,s^{-1}}} 
\newcommand\au{{\rm\,AU}} 
\newcommand\mpc{{\rm\,mpc}} 

\title{Tidal break-up of binary stars at the Galactic center and its consequences}

\author{Fabio Antonini}
\email{antonini@astro.rit.edu}
\affil{Department of Physics and  Center for Computational Relativity and Gravitation, Rochester Institute of Technology, 85 Lomb Memorial Drive, Rochester, NY 14623, USA }

\author{Joshua Faber}
\affil{Center for Computational Relativity and Gravitation, School of Mathematical Sciences, 
Rochester Institute of Technology, 85 Lomb Memorial Drive, Rochester, NY 14623, USA}

\author{Alessia Gualandris}
\author{David Merritt}
\affil{Department of Physics and Center for Computational Relativity and Gravitation, 
Rochester Institute of Technology, 85 Lomb Memorial
Drive, Rochester, NY 14623, USA}

\begin{abstract}
The tidal breakup of binary star systems by the supermassive black 
hole (SMBH) in the center of the galaxy has been suggested as the 
source of both the observed sample of hypervelocity stars (HVSs) in the 
halo of the Galaxy and the S-stars that remain in tight orbits around 
Sgr A*.  Here, we use a post-Newtonian N-body code to study 
the dynamics of main-sequence binaries on highly elliptical bound 
orbits whose periapses lie close to the SMBH, determining the 
properties of ejected and bound stars as well as collision products.  
Unlike previous studies, we follow binaries that remain bound for 
several revolutions around the SMBH, finding that 
in the case of relatively large periapses and highly inclined binaries
the Kozai resonance can lead to large periodic oscillations in the  
internal binary eccentricity and inclination.
Collisions and mergers of the binary elements are found to increase 
significantly for multiple orbits around the SMBH, while HVSs are 
primarily produced during a binary's first passage.  
This process can lead to stellar coalescence and eventually serve as
an important source of young stars at the Galactic center.
\end{abstract}

\subjectheadings{
black hole physics-Galaxy:center-Galaxy:kinematics and dynamics-stellar
dynamics}

\section{introduction}
The center of the Milky Way can be regarded as a rather unusual
 galactic nucleus in many respects.  
The supermassive black hole (SMBH) at the Galactic center 
has a mass of $M_{\rm MW}\sim 4\times 10^6\msun$
\citep{ghez08,gill09}, making it perhaps
the smallest SMBH with a well-determined mass \citep{ff2005}. 
Furthermore, while most known galactic nuclei have
relaxation times much longer than the Hubble time, the relatively
high density at the center of the Milky Way 
 ($\gap 10^6\,\rm stars\,pc^{-3}$) and the small mass of the SMBH
imply a two-body relaxation time that is as short as a few Gyr 
inside the SMBH influence radius \citep{alexander05}.
The combination of high stellar densities, large velocities, 
and short relaxation time effectively make
the inner parsec of the Milky Way a collisional system.

In a collisional environment, otherwise rare dynamical processes can
take place at an appreciable rate. These include dynamical encounters
between stars, binaries and higher-order systems.  If a massive black
hole is present, as is believed to be the case for most,
if not all, galactic nuclei, stars and binaries can interact with it
and be ejected with extreme velocities.  Gravitational encounters
involving stellar binaries and the SMBH have recently been studied by
several authors in the context of hypervelocity stars (HVSs) \citep{gps05,
brom06, GL:06, SE:07}. HVSs have such extreme velocities
($500-1000~{\rm \kms}$ as measured in the Galactic halo) that they require a
dynamical encounter with a massive black hole for their explanation
\citep{HI:88,YU:03}. The idea is that stellar binaries on low angular 
momentum orbits interact with the SMBH and are dissociated by its
strong tidal field. As a result, one component is captured by the SMBH
into a wide eccentric orbit while the other star is ejected to
infinity with large velocity.  After the fortunate discovery of the
first HVS in the Galactic halo \citep{b05}, a handful have been found
escaping the Galaxy at large speeds \citep{b06b}.
More recent surveys of hypervelocity
stars  found 16 stars with velocities 
larger than the Galactic escape velocity \citep{b07,b09}.

Given the high stellar density at the Galactic center, it is 
plausible that stars interacting with the SMBH will come  close enough to each
other that finite size effects become important and physical
collisions occur. In this case, the SMBH acts as catalyst for their
interaction.  \citet{GL:07} (hereafter GL07) first noted that the tidal breakup of a
stellar binary interacting with the SMBH can lead, at least for some orbital
parameters, to a physical collision soon afterwards.  If the two stars
collide with a relative impact speed smaller then the escape speed
from their surface, coalescence can occur, resulting in the formation of a
new, heavier star. 

Using Aarseth's direct integration scheme \citep{aarseth99} to
investigate the dynamics of the encounter, Ginsburg \& Loeb found that
the rate of collision events at the Galactic center is about ten times
smaller than the rate of formation of hypervelocity stars \citep[$
\sim 10^{-5} \rm yr^{-1}$,][]{b06b}, and that among the collision
products, stellar coalescence occurs in about twenty percent of the
cases.
In this paper we expand on their work and investigate the possibility
that stellar encounters with the SMBH are responsible for the
production of a population of rejuvenated stars around the SMBH.

If a binary  approaches the SMBH within its tidal disruption radius,
at the periapsis the stars can reach velocities a few percent of light speed, making relativistic effects potentially important
in determining the properties of the unbound population and the rate at which the stars are ejected at hypervelocities.
Here, we perform the first $N$-body simulations of binary-SMBH encounters that include post-Newtonian (PN) terms up to order 2.5,
for both  star-SMBH and star-star interactions.

In \S2 and  \S3, we describe the parameters we chose for our $N$-body evolutions 
and their physical motivation.
In \S4, we discuss the likelihood of tidal capture of stars by the SMBH 
during close passages of binary systems.  
\S5 and \S6 consider in greater detail the populations of HVSs and 
stellar collision/merger products, respectively, 
including mechanisms like Kozai resonance that are critical in 
understanding the long-term evolution of binary systems.   
A systematic comparison between Newtonian and PN integrations is 
provided throughout.  
Finally, in \S6, we provide a summary and discuss future avenues of study.

\section{Initial models and numerical methods}
We use $N$-body simulations to study the evolution of main sequence 
binaries as they make close passages by the SMBH at the Galactic center.
We determine the conditions under which gravitational interactions produce
HVSs, and the properties of the resulting
distribution of both unbound and bound stars.
All the simulations were carried out using the {\tt ARCHAIN}
\citep{MM:08} integrator, which includes PN
corrections to all pairwise forces up to order PN2.5. The code employs an
algorithmically regularized chain structure and the time-transformed
leapfrog scheme to accurately trace the motion of tight binaries with
arbitrarily large mass ratios \citep{MA:02,MM:06}.

Given the significant computational resources per run made necessary by the high precision of our simulations we
focus on a simplified set of initial conditions. 
We consider circular equal-mass binary systems with random initial orientations,
initial separations $a_0=0.05-0.2\au$, and individual stellar masses $M_*= 3-6\msun$. 
We define $M_{\rm b}$ to be the total mass of the binary system.  
Obviously, the choice of equal-mass components limits our work.
A treatment of binaries with a spectrum of masses similar for instance to that found 
in   \citet{DU:1991} is outside the scope of this paper, but could be included in a 
future work.

We give the binary a  tangential initial velocity $v_{\rm in}$ at a distance
$d=0.01-0.1$pc from the SMBH of mass
$M_{\bullet}=4\times 10^6\msun$, in effect setting the periapsis distance:
\begin{equation}
\label{per}
r_{\rm per}=\frac{\left( v_{\rm in}d \right)^2 }{2GM_{\bullet}-v_{\rm in}^2d}
\end{equation}
for a Newtonian elliptical orbit.  

Our initial binary separations  are close to the extremes of the interval
within which HVSs are produced.
Since the ejection velocity increases with the internal binary energy,
few stars will be ejected with velocities sufficiently large
to escape from the Galaxy for $a_0 > 0.2$AU, unless very large stellar masses are
considered \citep{gps05}. 
We also note that, for the initial distances $d$ we choose,
binaries with  $a_0 > 0.2$AU  are unlikely to survive for many orbits around the SMBH
due to close encounters with field stars \citep{PER:09}.
The lower limit of $0.05$AU is required by the consideration
that smaller separations will result in contact binaries.
GL07 pointed out that, 
assuming a constant probability per ln($a_0$) for $0.02<a_0<20$AU, 
 the probability to find a binary with a separation within the considered interval
is $\sim 20 \%$.
We place the binaries on highly elliptical orbits about the SMBH,
rather than e.g. on parabolic/hyperbolic orbits \citep{GQ:03}.

The origin of the young stars (and binaries) in the Galactic center is still
a matter of ongoing research.  One possibility is that stars/binaries
formed in-situ, i.e. at distances of a few tens of milliparsecs from
the SMBH. This model faces severe difficulties since tidal forces from
the SMBH inhibit star formation at these distances. Instead, it is typically assumed that the stars
need to form further away from the SMBH and then migrate to the center
within their lifetimes. Among the various migration models that have
been proposed, three produce orbits similar to those considered in this
work (i.e. highly eccentric): (i) the eccentric
disk instability scenario \citep{MAD:08}  
(ii) the cluster infall scenario aided
by an intermediate mass black hole (IMBH) \citep{GH:01} 
 and (iii) the triple disruption scenario  \citep{PER:09a}.  
The first model considers
a dynamical instability of an eccentric disk as the  mechanism that
drives the eccentricities of the stars away from their initial
values and produces near-radial orbits within a few Myr.  The
second model assumes  stars and binaries form during the collapse
of a giant molecular cloud and inspiral due to dynamical friction
toward the Galactic center. The cluster, which is subject to tidal disruption,
can reach distances of $10-50\,\rm mpc$ if it harbors an IMBH of mass
$\gap10^3\msun$ at its center.
Tidally removed stars can then be scattered onto eccentric orbits
by the IMBH \citep{MA:09}.
Finally, in the  triple disruption scenario, young and runaway stars can
 form by disruptions of triples by the SMBH, which in some cases
results in the capture of binaries on close orbits.

At least some of the binary progenitors of observed HVSs were likely to 
have been part of the young stellar disk observed at the Galactic center,
whose inner edge lies at a radius $0.1$pc from the SMBH \citep{LBK},
and this fact motivated our choice for the maximum value of $d$.
The distance $0.01$pc adopted as a minimum value for $d$ 
is comparable to the radii of the smallest observed
stellar orbits at the Galactic center \citep{EC:97,GE:97}.

\section{Basic relations}

Three radii play a fundamental role in our simulations:
\begin{itemize}
\item[$\bullet$]the periapsis separation $r_{\rm per}$  of the initial orbit of the binary with respect to the SMBH;
\item[$\bullet$]the tidal disruption radius $r_{\rm t}$ of a single star due to the SMBH; 
\item[$\bullet$]the tidal disruption (i.e. breakup)
radius $r_{\rm bt}$ of the binary  due to the SMBH.
\end{itemize}
A binary with initial separation $a_0$ will be broken apart by tidal
forces if its center of mass approaches the massive object within the
distance \citep{ML:05, SE:09}
\begin{equation}\label{rbt}
r_{bt} \sim \left( 3 \frac{M_{\bullet}}{M_{\rm b}} \right)^{1/3}a_0. 
\end{equation}

As Hills (1988) noted, the tidal disruption of single stars
is an important issue in the context of binaries. 
Because of their low densities,
main sequence (MS) stars do not survive close interactions with the SMBH; 
small initial
velocities take the stars too close to the SMBH and no HVSs are
produced (compact objects, e.g., neutron stars and
white dwarfs, can survive significantly closer approaches).  
Tidal disruption occurs for stars that approach the SMBH
more closely than $r_t$, where
\begin{equation}\label{rt}
r_{\rm t} \sim \left( \frac{M_{\bullet}}{M_*} \right)^{1/3} R_*,
\end{equation}
with $R_*$ the stellar radius.  We assume a mass-radius relation
$R_*/R_{\odot}=\left( M_*/M_{\odot} \right)^{0.75}$ \citep{HA:04}
which yields $R_*=0.01~\rm{AU}$ for $M_* = 3~\rm{M_{\odot}}$, and
$R_*=0.016~\rm{AU}$ for $M_* = 6~\rm{M_{\odot}}$.  Stars with orbits
meeting the criterion $r_{\rm per}<r_t$ are fully disrupted \citep{LC:86,
  EK:89, FA:05}.  Note that this condition is stronger than the
classical Roche limit formula given by \citet{pac:71} and
\citet{egg:83}. Even when tidal disruption does not occur, we expect
that mass will be
stripped from the outer regions of any star passing within the Roche limit, which would have a fundamental impact on the
subsequent evolution and orbit of the stars.

For $M_* = 3\msun$ and $M_{\bullet}=M_{\rm MW}$, $r_{t}$ is $\sim 1.10~\rm{AU}$, while for $M_* =
6\msun$ the corresponding value of $r_{\rm t}$ is $ \sim
1.40~\rm{AU}$ (i.e., MS stars with larger masses are
disrupted at larger distances).

A comparison of equations~(\ref{rbt}) and (\ref{rt}) reveals that for MS
 binaries we always have $r_{\rm bt}>r_{\rm t}$ ( {\rm for contact binaries instead
 it is possible to disrupt a star before  ejection occurs}).
Our numerical
simulations demonstrate that  for large binary separations where
$r_{\rm per}<r_{\rm bt}$, the two stars remain bound to each other up to
distances $\sim r_{\rm bt}/2$. This can be understood if the dynamical
crossing time of the binary at $r_{\rm bt}$ is of the order of the time
scale of tidal breakup of the binary.  
Furthermore even at later times, after the
binary is broken apart, both stars remain close to the original
Keplerian orbit of the binary around the SMBH until reaching
distances $\sim r_{\rm per}$.  We will come back to this point below.

Using equations (\ref{per}) and (\ref{rt}) we can roughly estimate 
the lowest value of $v_{\rm in}$ that avoids tidal disruption of the
individual stars.  The condition $r_{\rm per}> r_{\rm t}$ yields
\begin{equation}
\label{vc1} 
v_{\rm in} \ge v^c_{\rm 1} \approx 
\left[\frac{2GM_{\bullet}}{\left(d^2/r_{\rm t}+ d \right)}\right]^{1/2}.
\end{equation}
For $d=0.01(0.1)$pc, this critical velocity is 
$v^c_{\rm 1}=44(4.0)\kms$ for $M_*=3\msun$ and 
$50(4.5)\kms$ for $M_*= 6\msun$.  
When PN corrections are included the orbits are
no longer Keplerian and the periastron distance can be found by
including relativistic terms in the expression for the potential
\citep{Soffel:89}:
\begin{equation}\label{RP}
V(r)=-\frac{GM_{\bullet}M_{\rm b}}{r}+\frac{L^{2}}{2M_{\rm b} r^{2}}-\frac{GM_{\bullet} L^{2}}{c^{2}M_{\rm b}r^{3}}, 
\end{equation}
where $L$ is the binary angular momentum and $c$ the speed of light.
For a given initial velocity, the net relativistic effect consists of
a decrease of $r_{\rm per}$, yielding a higher limit for the initial
velocity than for the Newtonian case. This value, in the case $d=0.01$pc, is $v^c_{\rm 1}
\sim 47$ and $53\kms$ for $M_*=3$ and $6\msun$ respectively.  For
$v_{\rm in}<v^c_{\rm 1}$ stars would be tidally disrupted.

We can derive an expression for the maximum initial velocity that
should still be able to produce a HVS.  The minimum condition is that
the binary itself is broken apart.  Together with equation~(\ref{rbt}), this
gives
\begin{equation} \label{vc2}
v_{\rm in} \le v^c_{\rm 2} \approx \left[\frac{2GM_{\bullet}}{\left(d^2/r_{\rm bt}+d \right)}\right]^{1/2}.
\end{equation} 
Very few HVSs are likely to be ejected at $v_{\rm in}>v^c_{\rm 2}$. 

In this approximate model, HVSs are produced for
$v^c_{\rm 1}<v_{\rm in}<v^c_{\rm 2}$. Between these limits, one member of  a binary may be
captured by the SMBH and begin to orbit it at large eccentricity,
while the other star is ejected with a velocity $v_{\rm ej}$ that is
larger than the escape velocity from the Galaxy.  An approximate
expression for $v_{\rm ej}$ was obtained by \citet{HI:88} and
\citet{YU:03}:
\begin{eqnarray} \label{VEJ}
v_{\rm ej} \approx 1770 \left(\frac{a_0}{0.1AU}\right)^{-1/2}  \left(\frac{M_{\rm b}}{2M_{\odot}} \right)^{1/3}  \nonumber\\
  \left(\frac{M_{\bullet}}{3.5\times 10^{6}M_{\odot}} \right)^{1/6} 
f_{\rm R}~{\rm km~s}^{-1} ~~~
\end{eqnarray} 
with $f_{\rm R}$ a function of the dimensionless closest approach parameter $D=(r_{\rm per}/a_0)[2~M_{\bullet}/10^6~M_{\rm b}]^{-1/3}$  .
An expression for $f_{\rm R}$ was derived by \citet{brom06}:
\begin{eqnarray} \label{fr}
f_{\rm R}=0.774+(0.0204+[-6.23\times10^{-4}+\{7.62\times10^{-6}+  \nonumber \\
 (-4.24\times10^{-8}+8.62\times10^{-11}D)D\}D]D)D, ~~~~~  
\end{eqnarray}
which reproduces the spectrum of ejection velocities for binaries   initially unbound relative to the SMBH.
For bound orbits, the ejection speeds will typically be smaller, unless very large apoapses are adopted.

\begin{figure*}[ht!]
\begin{center}
$\begin{array}{cc}
\includegraphics[angle=0,width=3.5in]{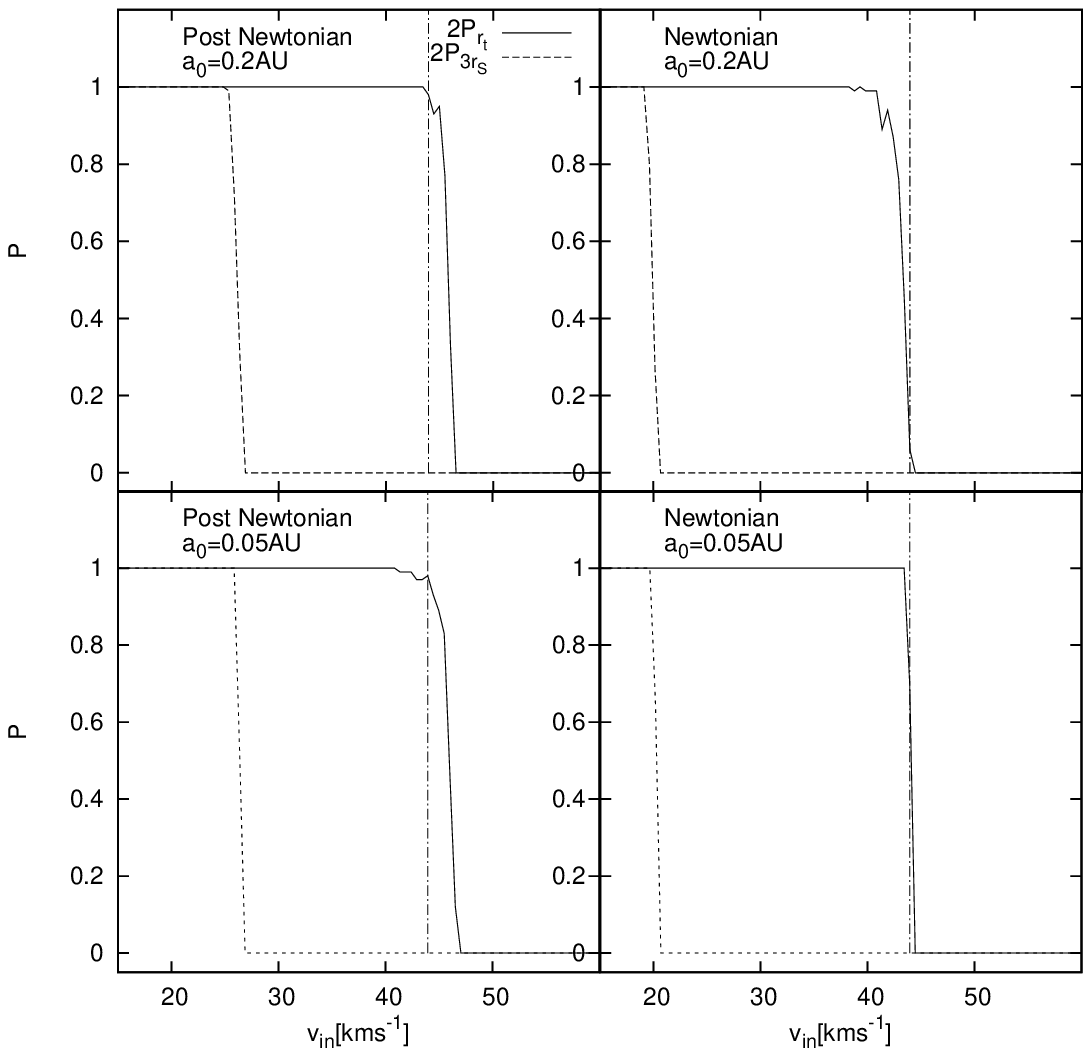} &
\includegraphics[angle=0,width=3.5in]{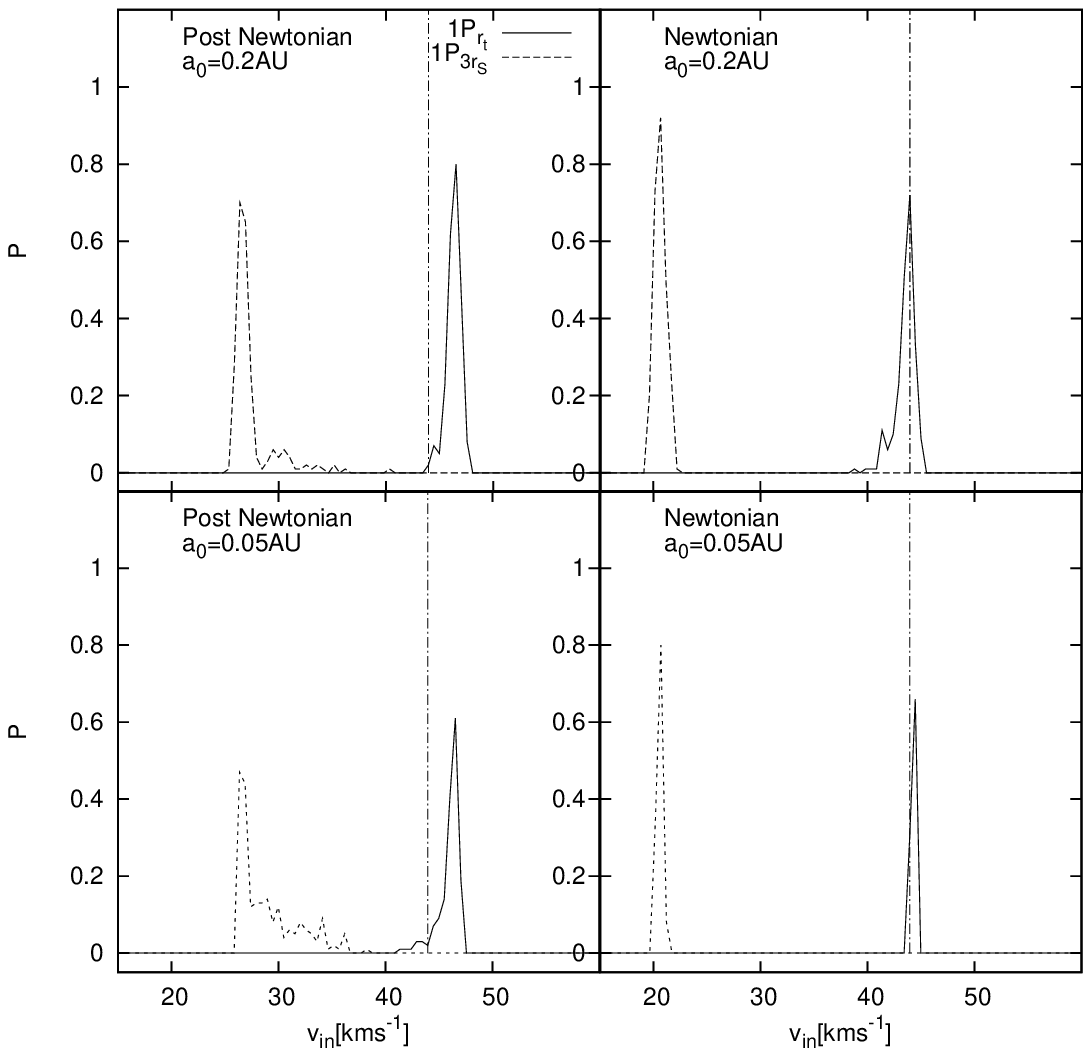} 
\end{array}$
\caption{Probability that both stars (2P in the left panels) 
or only one star (1P in the right panels) of the
  binary fall within the tidal disruption radius $r_{\rm t}$ (solid lines) and the
  SMBH ISCO $3R_{\rm S}$ (dashed lines) as a function of the initial
velocity $v_{\rm in}$.  The  vertical dot-dashed line represents the 
   value of $v_{\rm 1}^c$.}
\label{fig1}
\end{center}
\end{figure*}

\section{Tidal captures}\label{sec:tidalcapture}

We first devised a set of simulations to determine the rate at
which stars are captured by the SMBH.  
Here, ``capture'' means that a star passes either inside the
stellar tidal disruption radius $r_{\rm t}$, or inside 
the innermost stable circular orbit (ISCO) 
at $3R_{\rm S}$, $R_{\rm S}\equiv 2GM_\bullet/c^2$; $R_{\rm ISCO}\approx 0.25\au$ for 
 $M_\bullet = 4 \times 10^6\msun$.
Naively, one expects that passage of the binary within either
radius would result in both stars being captured.
In fact, due to the finite size of the binary, it is possible for one
star to be destroyed and the other to escape as a HVS.

We performed $\sim36000$
integrations with equal mass $M_* = 3\msun$ binaries, $a_0=0.05$ or
$0.2\au$ and $d=0.01$pc, choosing random orientations and initial tangential
velocities in the range $4\kms \le v_{\rm in} \le 85\kms$.  
For each of the two binary
separations, we considered both Newtonian and PN cases.
Stars were treated as point masses, and we recorded all instances 
when a star fell inside $3R_{\rm S}$ or  $r_{\rm t}$.  
Given the possibility that only one of the two stars is captured
during a close passage to the SMBH, it is important to include
initial conditions with $r_{\rm per}\lesssim r_{\rm t}$ 
(or equivalently $v_{\rm in}<v^c_{\rm 1}$).

Figure~\ref{fig1} shows the probability that one or both stars
will pass inside $3R_{\rm S}$ and/or $r_{\rm t}$ as a function of the 
initial velocity $v_{\rm in}$.
As a rule, passage of one star inside $r_{\rm t}$ or $3R_{\rm S}$ implies
that both stars are captured.

The probability that both stars are tidally disrupted rises
essentially to one at $v_{\rm in}\lap v_{\rm 1}^c \approx 44\kms$
in the classical case, and at a slightly higher velocity
in the relativistic case.

The relativistic shift is well known:
for a Keplerian orbit with $r_{\rm per}\sim 1\au$,
the difference in the relativistic periapsis with respect to the
Newtonian value is $\sim 0.12\au$, attributable to the attractive third
term that appears in the relativistic potential in equation (\ref{RP}).  
Thus, when the binary has $v_{\rm in}< 50\kms$, both stars are likely to be tidally
disrupted by the SMBH.
 
  \citet{GL:06} noted that
  the typical impact parameter
  that leads to the break-up of the binary 
  by the black hole, is much larger than $r_t$,
  the tidal disruption radius for a single star.
  On this basis, they ignored stellar tidal disruptions.
  Our initial conditions are essentially the same as theirs;
  but as shown in Figure~\ref{fig1},  these initial conditions would 
  result in stellar tidal  disruption for a large fraction of the orbits,
  contrary to  the assumption of Ginsburg \& Loeb.
  In their subsequent paper,  \citet{GL:07}   assumed  
  $v_{in} < 25 \kms $,  which {\it always} leads to stellar disruption.
Our study demonstrates that the relevant distance in the problem is 
  the orbital periapsis of the center of mass of the binary (equation 
  \ref{per}), which is also the distance of the closest approach
  of the stars to the SMBH.  Indeed, for   a  wide range of initial velocities,
  the stars  penetrate deeply   the SMBH's  potential well.
  For    $v_{\rm in} \gtrsim 45 \kms $ tidal disruption is avoided but
   tidal perturbations   of the SMBH would still be expected to have a 
  significant influence on the subsequent evolution of the  stars.

Figure~\ref{fig1} also indicates that it is possible for only one star to 
be captured.
In our calculations this mechanism does not produce a significant number of HVSs 
since this occurs for a narrow range
of initial velocities: $v_{\rm in}\approx v_{\rm 1}^c$ in the classical case,
and $v_{\rm in}\gap v_{\rm 1}^c$ in the relativistic case.
Between these cases we found that for wider binaries ($a_0=0.2{\rm AU}$)
the non-disrupted star is typically still on a bound orbit around the SMBH while
for tighter binaries ($a_0=0.05{\rm AU}$)  the ejection probability is around $\sim 80 \%$
and the mean ejection velocity is $\sim 3000 \kms$.

If $r_{\rm per}$ is slightly larger than $r_{\rm t}$,
the binary is  usually disrupted at the first encounter with the SMBH.
However we found a small fraction of orbits for which the binary survived 
for longer times.  
In our simulations, unbinding
of the binary ($E_{\rm b} \approx 0$) occurs approximately at $2\au$ for $a_0
\approx 0.05 ~\rm{AU}$ and at $\sim 10\au$ for $a_0 \approx 0.2\au$.
In the subsequent evolution the separation between the
stars does not change appreciably until the periapsis passage.  After
this point, the stars can either separate or become bound to each
other again, continuing to orbit the SMBH as a binary.  When this
occurs, the minimum separation between the stars can become very small
at later times.  

\begin{figure*}
\begin{center}
\includegraphics[angle=270,width=5.7in]{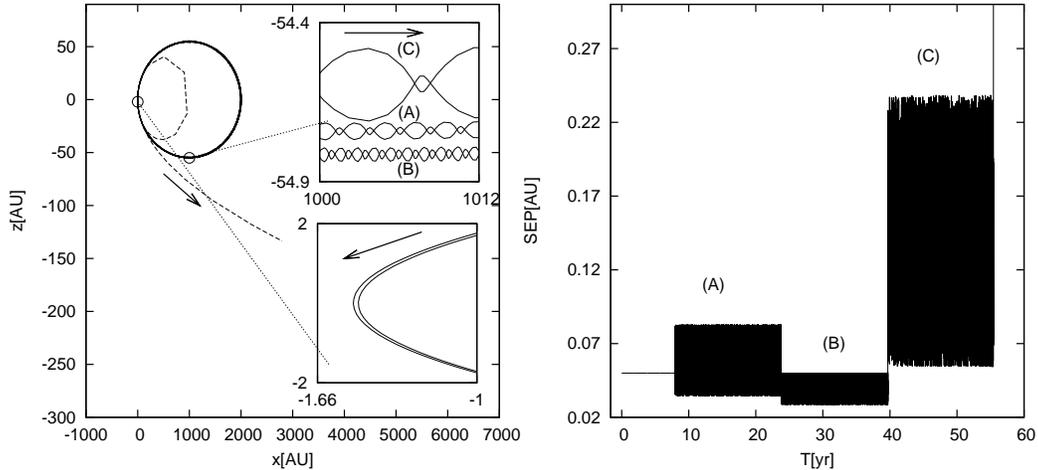}
\caption{Example of an orbit whose periastron separation from the SMBH is
  $r_{\rm per}\gtrsim r_t$. The binary is initially located at $x=2000 {\rm AU}(0.01{\rm pc}) $
  with a purely tangential initial velocity of $v_z=51\kms$.
  The stellar masses are $M_*=3\msun$ with initial binary separation 
  $a_0=0.05~\rm{AU}$. The left panel shows the orbit of the stars.
  The binary makes three revolutions around the SMBH before it is
  disrupted, after this point the individual stellar orbits are given by dashed lines.
  The upper inset panel schematically displays the different internal
  binary orbits during the first (A), second (B), and final loop (C).
  The lower inset panel shows the stars' orbits with respect to the SMBH
  during the first encounter.  Note that the x and z-axes have
  very different scales.  The right panel displays the change of the
  separation between the stars with time. The upper and lower limits
  of each ``block'' represent apoapsis and periapsis of the internal
  binary orbit respectively.}
\label{fig2}
\end{center}
\end{figure*}
An example of the latter situation is illustrated in
Figure~\ref{fig2}, which shows the evolution of the separation
between the stars, as well as the orbit of the binary with respect to
the SMBH, for an initial separation $a_0 \sim 0.05\au$ and
initial velocity $v_{\rm in} = 51\kms$.  After the first encounter with
the SMBH, where the internal binary eccentricity suddenly changes, the
periapsis separation between the stars is $\sim 0.035\au$ while the
apoapsis separation is $\sim 0.08\au$.  After the second encounter,
the stars orbit each other on an elliptical orbit with periapsis
separation $\sim 0.028\au$.  The third encounter with the massive
object results in a very eccentric orbit.  Finally, after four
encounters with the SMBH the binary is disrupted and the two stars
begin to orbit the SMBH on two different high eccentricity orbits.

The previous arguments suggest that the eventual shrinkage of the
orbit after the encounter with the SMBH can easily trigger mass
transfer and, in extreme cases, a coalescence between the stars.  The
merger remnant would orbit the SMBH on a highly eccentric orbit.
In the case displayed in Figure~\ref{fig2}, the minimum separation
between the stars is small enough to allow mass transfer after the
second passage close to the SMBH.  The Roche limit corresponding to
the onset of mass transfer in binary systems of equal-mass stars would
be $\sim 2.8 R_*$ while the dynamical instability limit in the case of
co-rotating equal-mass MS stars is located around $\sim 2.4 R_*$
\citep{RS:92,RS:95}; for the case of eccentric binaries the radius of
the Roche lobe is similar to that of circular binaries when calculated
at periastron \citep{RS:05}. We also stress that this scenario would
hold for higher values of $r_{\rm per}$, in which case it would not be
restricted to small periastron separations, as will be shown below.

From this first set of integrations we deduce that cases in which
only a single star is captured are relatively rare, and occur when
$r_{\rm per}\approx r_{\rm t}$.
Stars ejected from these  orbits are expected to contribute weakly to 
the population of HVSs.
This conclusion should hold in general when different values of  
$d$ and $M_*$ are chosen.

\begin{table}
\caption{Initial parameters. \label{t1}} 
\begin{tabular}{lllll}
 \hline
 $a_0$[AU]  & $M_*$[M$_\odot$] & $d$[pc] &  Gravity   \\ 
\hline
 $0.2$      & $3$ & $0.01$  & post-Newtonian and Newtonian & \\
 $0.2$      & $6$ & $0.01$  & post-Newtonian and Newtonian & \\
 $0.05$     & $3$ & $0.01$  & post-Newtonian and Newtonian & \\
 $0.05$     & $6$ & $0.01$  & post-Newtonian and Newtonian &\\
 $0.2$      & $3$ & $0.1$ & post-Newtonian &\\
 $0.2$      & $6$ & $0.1$ & post-Newtonian &\\
 $0.05$     & $3$ & $0.1$ & post-Newtonian &\\
 $0.05$     & $6$ & $0.1$ & post-Newtonian &\\
\hline
\end{tabular}
\end{table}

\begin{table*}
 \centering
\caption{Ejection probabilities for the post-Newtonian (Newtonian) runs described in Table~\protect\ref{t1} for cases where $r_{\rm per}<r_{\rm bt}$. \label{t2}} 
\begin{tabular}{lllllll}
  \hline
 $a_0$[AU]  & $M_*$[M$_\odot$] & $d$[pc] & Escape with                       & Escape with                & Escape with                  & Escape SMBH   \\
\phantom   & \phantom            & \phantom& $v_{\rm ej} >800 \kms$    &$v_{\rm ej} >1000 \kms$ &  $v_{\rm ej} >1400 \kms$  & \\ 
 \hline
 $0.2$      & $3$ & $0.01$   &  11.0(9.57)  & 9.71 (7.35)         &   6.60   (7.02)         & 28.2 (24.3) \\
 $0.2$      & $6$ & $0.01$   &  55.5(56.3)  & 53.5 (49.7)         &   27.7   (27.5)         & 57.3 (60.7) \\
 $0.05$    & $3$ & $0.01$   &  68.6(62.0)  & 66.7 (62.0)         &   65.6   (62.0)         & 72.0 (62.0) \\
 $0.05$    & $6$ & $0.01$   &  54.2(66.0)  & 52.5  (65.1)        &   52.5   (65.1)         & 65.5 (66.3) \\
 $0.2$      & $3$ & $0.1$     &  80.0           & 74.5                   &   55.5                     & 89.1           \\
 $0.2$      & $6$ & $0.1$     &  80.9           & 79.1                   &   71.8                     & 85.5           \\
 $0.05$    & $3$ & $0.1$     &  69.3           & 69.3                   &   69.3                     & 71.6 \\
 $0.05$    & $6$ & $0.1$     &  44.3           & 44.3                   &   44.3                     & 44.3 \\
 \hline
\end{tabular}
\end{table*}

\section{Ejection of HVS}

For the remainder of this paper, we focus on cases where $r_{\rm per} > r_{\rm t}$.
This allows us to neglect possible tidal effects on stars that would
require a hydrodynamic treatment.

The tidal disruption of binaries by $\rm Sgr A^*$ is generally
accepted to be the main source of HVSs (stars ejected with  $v>1000 \kms$) in our galaxy.
The maximum velocity that can be achieved with the classical
binary supernova scenario \citep{BL:61} is $\lesssim 300\kms$
for $3\msun$ stars. Interactions with a massive compact object
seem necessary in order to explain the extreme velocities
of HVSs.

In this section we explore this mechanism by integrating a set of  $\sim10000$
orbits featuring binary stars closely interacting with a SMBH.
Two different stellar masses were considered:
$M_*=3\msun$ and $6\msun$. The runs
were performed for initial velocities that correspond to periapsis 
between $r_{\rm t}$ and $6r_{\rm bt}$.  In all the simulations the final
integration time was fixed to $60$ orbital periods of the initial binary
orbit around the SMBH.  For the adopted periapsis distances, this integration time 
is long enough for the perturbations induced by the SMBH 
on the internal binary's orbit to grow dramatically.
Table~\ref{t1} summarizes the initial parameters we chose for 
the integrations. 

Because the binary is initially bound to the
SMBH, no HVS is produced when two stars merge (i.e.,
if the relative velocity is lower than the escape velocity from
the stellar surface after a collision). On the other hand, if the stars collide without
merging the ejection of a single star is still possible. Hence, in the
following, we assume that a HVS ejection is possible even after the
stars collide, unless the final product is a merger.

\begin{figure*}[ht!]
\begin{center}
\includegraphics[angle=270,width=5.6in]{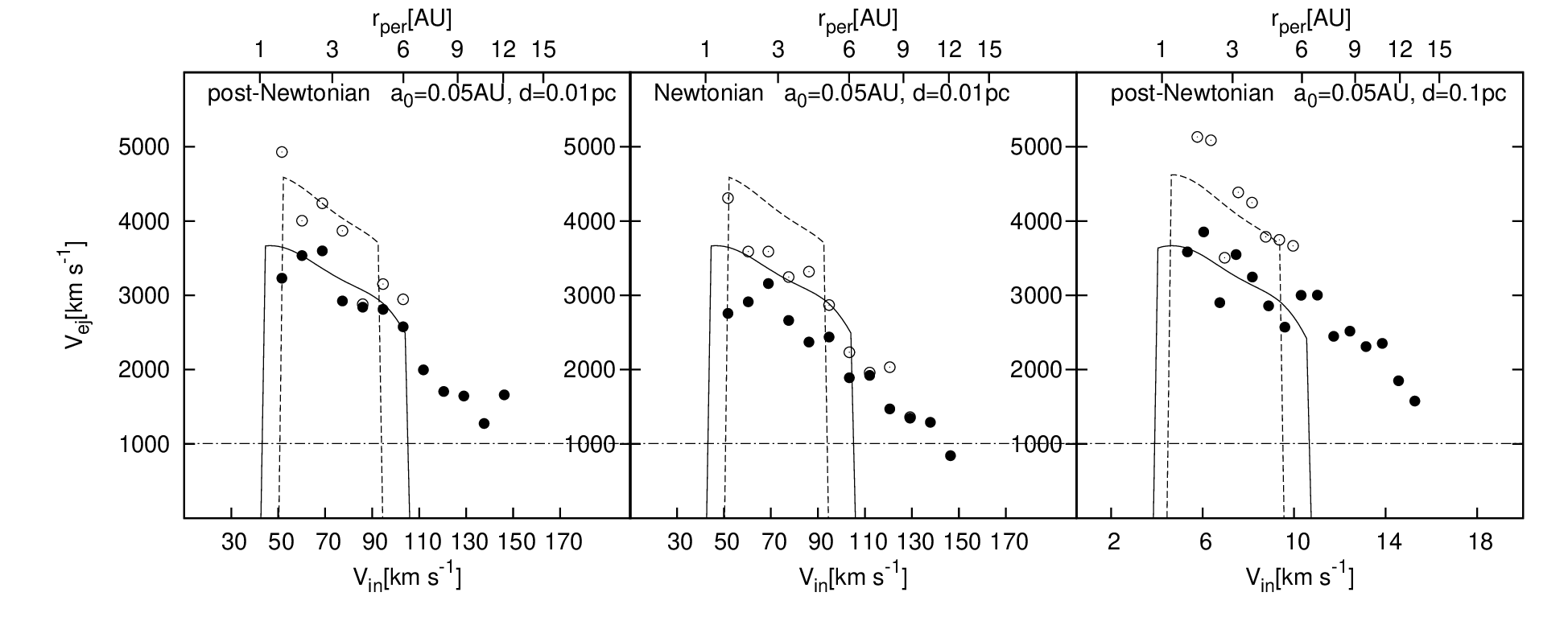} 
\includegraphics[angle=270,width=5.6in]{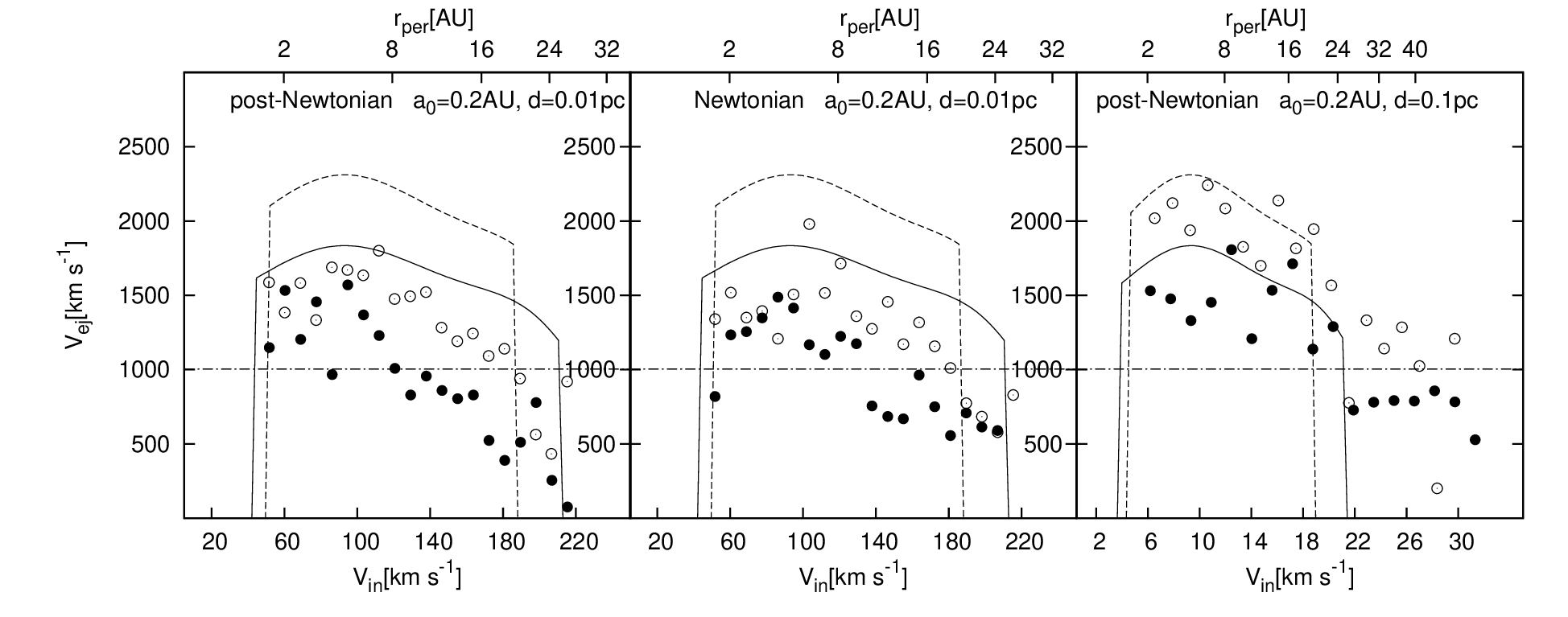} 
\caption{Average ejection velocity of the stars unbound to the SMBH
 for stellar masses $3~\rm{M_{\odot}}$ (filled circles) and
  $6~\rm{M_{\odot}}$(open circles).  The  model described in \S3
 is superposed using dashed lines for $6~\rm{M_{\odot}}$
and solid lines for  $3~\rm{M_{\odot}}$. 
The horizontal dot-dashed  line shows the
  approximate escape velocity from the Milky Way ($\sim1000\kms$).}
\label{fig3}
\end{center}
\end{figure*}

\subsection{Unbound population}

Figure~\ref{fig3} displays the average asymptotic ejection velocity of
the stars unbound from the central object, as well as a comparison
with the approximate model described by equations (\ref{vc1}), 
(\ref{vc2}) and (\ref{VEJ}).  
Our results are in fair
agreement with the predicted model,
at least for $d=0.1\rm{pc}$.
When the apoapsis  of the initial orbit is reduced to $d=0.01\rm{pc}$ and $a_0=0.2AU$,
the ejection speed of the unbound stars can be a factor of $~2$ smaller than the typical values found for
initially unbound stars.
For  $a_0=0.05$AU, on the other hand, the results are  in good agreement with the model. 
The larger discrepancy in the case of wider binaries 
 with respect to the theoretical model (i.e., unbound orbits )
is due to their larger value of $r_{\rm bt}$.
Since the ejection velocity  depends strongly
on the Keplerian velocity of the binary when its
components start to separate, the contribution 
of the binary orbital energy  to $v_{\rm ej}$
will be more important when 
the disruption occurs at larger distances from the SMBH.
 Another  source of the observed discrepancy  is due to the artificial truncation done at $v^c_{\rm 2}$, beyond which a number of unbound stars are still produced which represent the tail of the
distribution with the lowest values of $v_{\rm ej}$. 
These cases are mainly due to binaries which are broken 
apart after few orbits around the central object. 
At each encounter the internal eccentricity of the binary changes, 
and there is a new chance for the stars to be separated
and for a member to be ejected.

 \begin{figure}
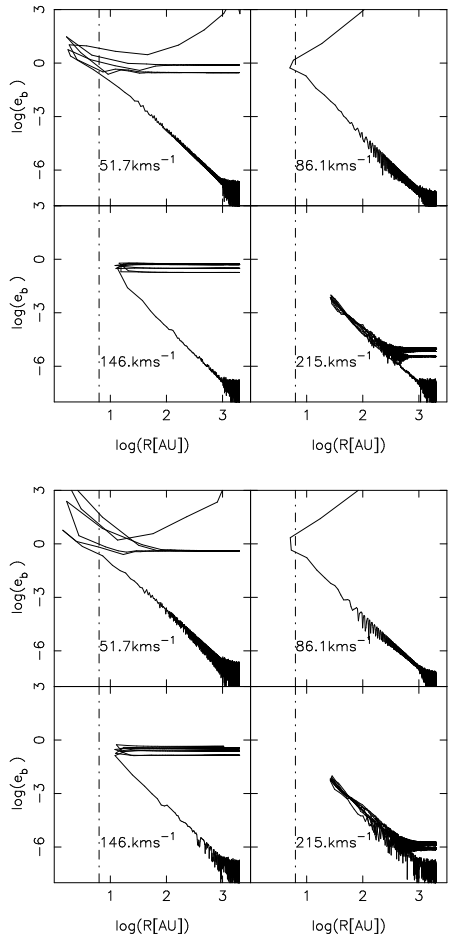

\begin{center}
$\begin{array}{cc} 
\includegraphics[angle=0,width=2.3in]{fig4a.ps} \\
\\
\includegraphics[angle=0,width=2.3in]{fig4b.ps} 
\end{array}$
\caption{Internal binary eccentricity $e_{\rm b}$ as a function of the distance
  $R$ to the SMBH for different initial velocities (displayed in the panels). The binaries have initial
  separation $a_0=0.05\au$, stellar masses  $3\msun$ and the apoapsis is $0.01$pc.
  Top: Newtonian integrations. Bottom: PN integrations.
The vertical dot-dashed line gives the value of $r_{\rm bt}$.
For small $v_{\rm in}$,  the binary approaches very near the SMBH  and the encounter
results in the tidal break-up of the binary.  In some cases 
the binary can complete several orbits around the SMBH before its stellar components are finally separated, even though $r_{\rm per}<r_{\rm bt}$ . 
Binaries with larger periapsis separations, or, equivalently, larger initial velocities,
are likely to survive for a larger number of orbits.
During each periapsis passage, gravitational interactions produce significant changes
in the internal eccentricity of the binary.}
\label{fig4}
\end{center}
\end{figure}

To demonstrate the orbital evolution during an integration, 
Figure~\ref{fig4} shows the change of the internal binary
eccentricity $e_{\rm b}$ as a function of the distance to the SMBH for several representative cases 
with different initial velocities in both PN and Newtonian
gravity. The eccentricity was evaluated from the specific angular momentum and
binding energy of the binary: $e=\sqrt{1+2EL^2/(GM_{\rm b})^2}$.
The initial separations, stellar masses, and apoapses were given respectively by $a_0=0.05\au$,   $M_*=3\msun$ and $d=0.01$pc.
The eccentricity behaves similarly for
different initial conditions, increasing suddenly at $\sim r_{\rm bt}$,
though in some cases the orbits may do several passages within $r_{\rm bt}$
while maintaining a negative internal binary energy.  
When the binary has a small periapsis, several transitions between 
 $e_{\rm b} > 1$ (unbound)  and  $< 1$ (bound) are possible.

Results from our runs whose orbits have periapses within the binary tidal disruption radius
(i.e., $r_{\rm per}<r_{\rm bt}$) are described in Table~\ref{t2}.  In general, we find for wider binaries with $a_0=0.2$AU, 
lighter binaries with $M_*=3\msun$ 
are less likely to produce HVSs capable of escaping the galaxy than heavier binaries with $M_*=6\msun$, though they
can produce a significant population of stars bound to the galaxy but unbound from the central SMBH. 
For tighter binaries with $a_0=0.05$AU the situation is reversed, and lighter binaries are more likely to 
eject HVSs from the galaxy.  For these tighter binaries, there are very few ejections of stars that unbind
from the SMBH but remain bound to the galaxy, regardless of the stellar mass.

A comparison between
the velocity distributions of the ejected stars in the Newtonian and
PN cases reveals that relativistic effects  have
a slight influence on the mean properties of the sample.
In particular for $a_0=0.05$AU and $d=0.01$pc the mean ejection speed
is larger when the relativistic corrections are included. 
For larger separations, instead, this effect is absent and the 
simulations show similar results. 
However, we were not able to identify a clear systematic effect due to the PN corrections 
in the properties of the ejected stars.

\subsection{The Bound Population}
Each star ejected during a binary-SMBH encounter is associated with a
captured companion that loses energy in the process and becomes more
tightly bound to the SMBH on a high eccentricity ($e\gap 0.9$) orbit.
Bound stars are also produced after the tidal break up of a binary if
neither star is ejected.
The orbital parameters of the bound stars will be strongly correlated
with the amount of energy carried away by the companion.
For an initially unbound  orbit the apoapsis
distance of the bound star is approximately
\begin{equation} \label{ap}
a \sim \frac{GM_{\bullet}}{v_{\rm ej}^2},
\end{equation}
and the corresponding orbital period is 
\begin{equation}
P \sim \frac{GM_{\bullet}}{v_{\rm ej}^3}.
\end{equation}  
The mean value of the semimajor axis $A$ (with respect to the
massive body) for three-body exchanges with equal-mass binaries 
can be expressed as \citep{HI:91}
\begin{equation}
\langle A\rangle \approx 0.56 \left( \frac{M_{\bullet}}{M_{\rm b}} \right)^{2/3}a_0.
\end{equation}

Bound stars can also be produced when the binary components
merge.  A coalescence remnant will not be able to escape the
SMBH gravitational potential if the initial binary is bound to the
SMBH unless significant mass loss occurs during coalescence.

\begin{figure}
\begin{center}
$\begin{array}{cc}
\includegraphics[angle=270,width=1.7in]{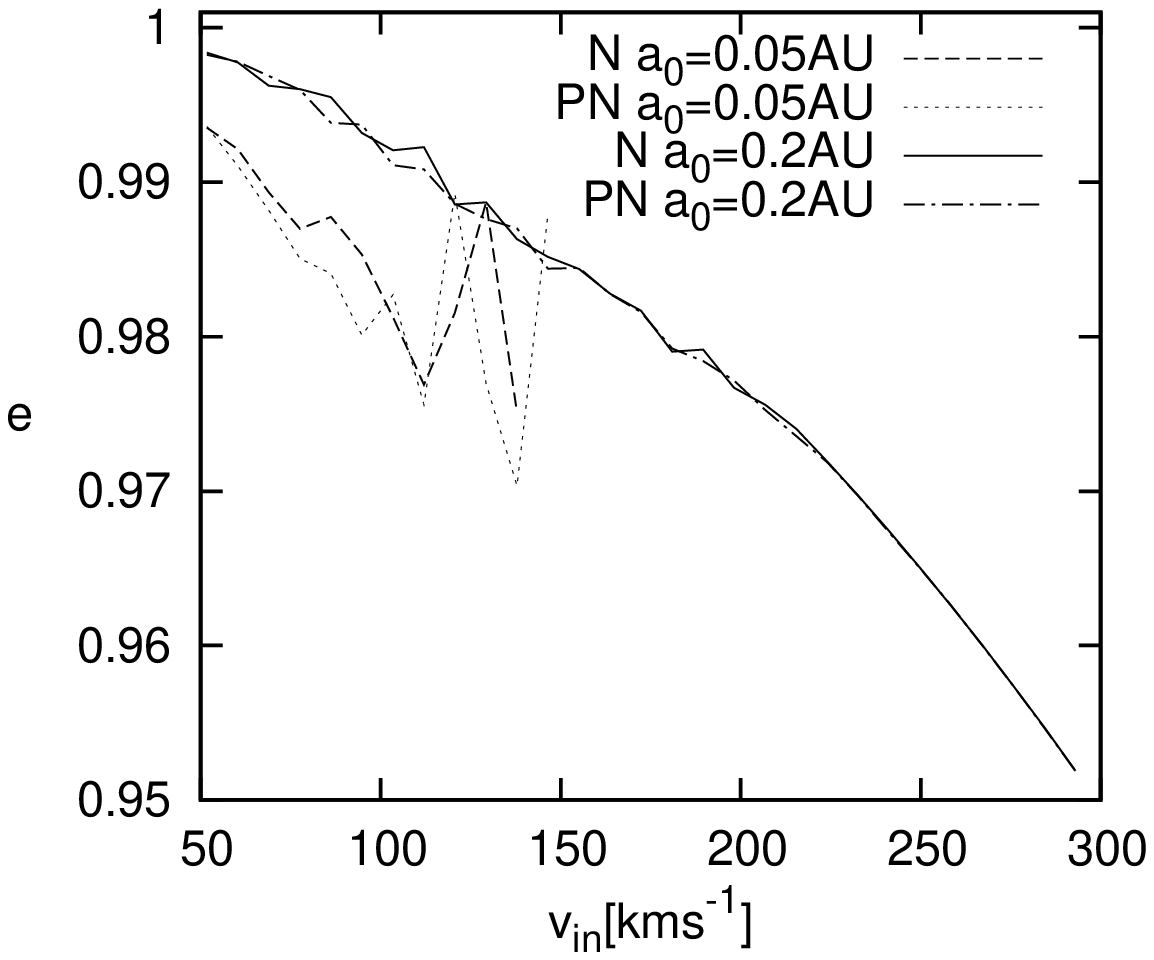}&
\includegraphics[angle=270,width=1.7in]{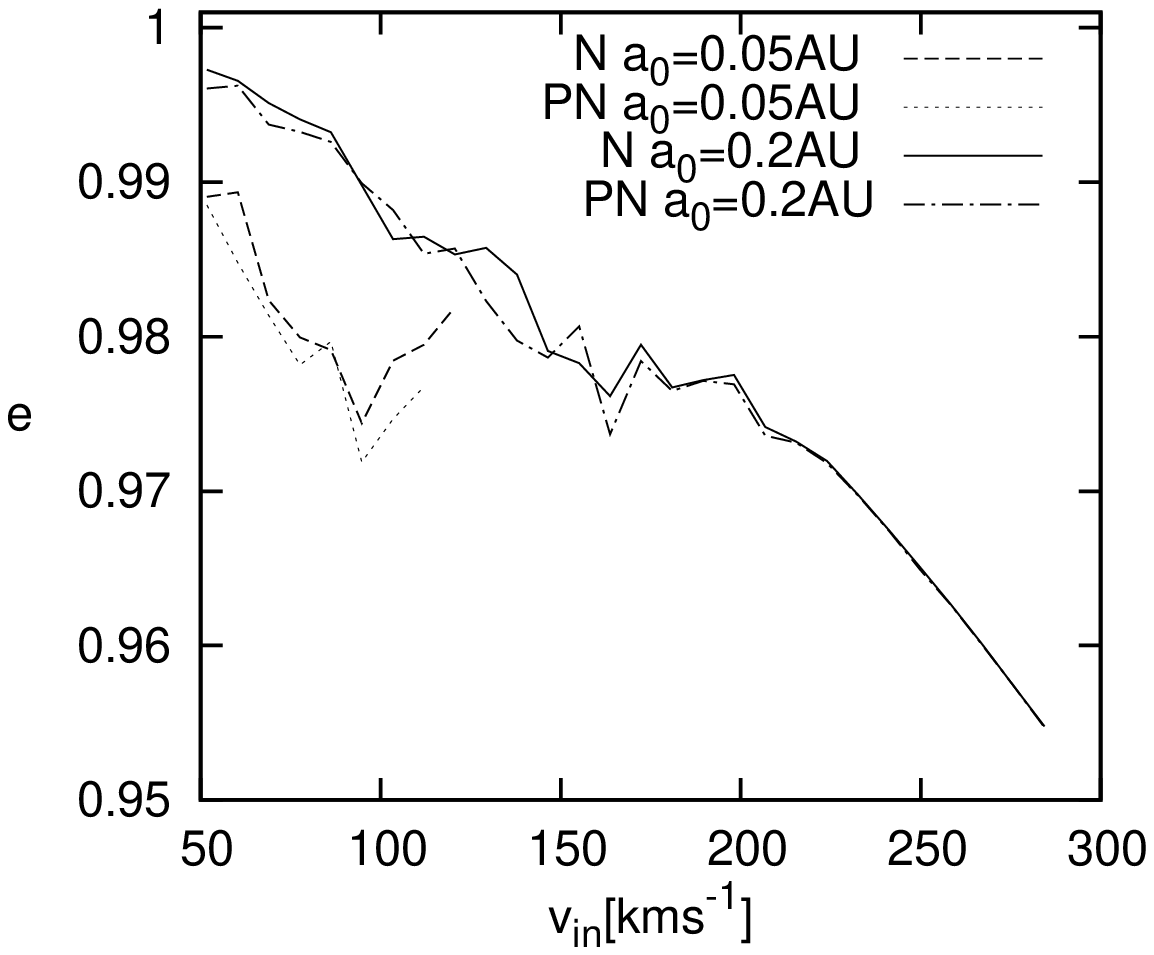}\\
\includegraphics[angle=270,width=1.7in]{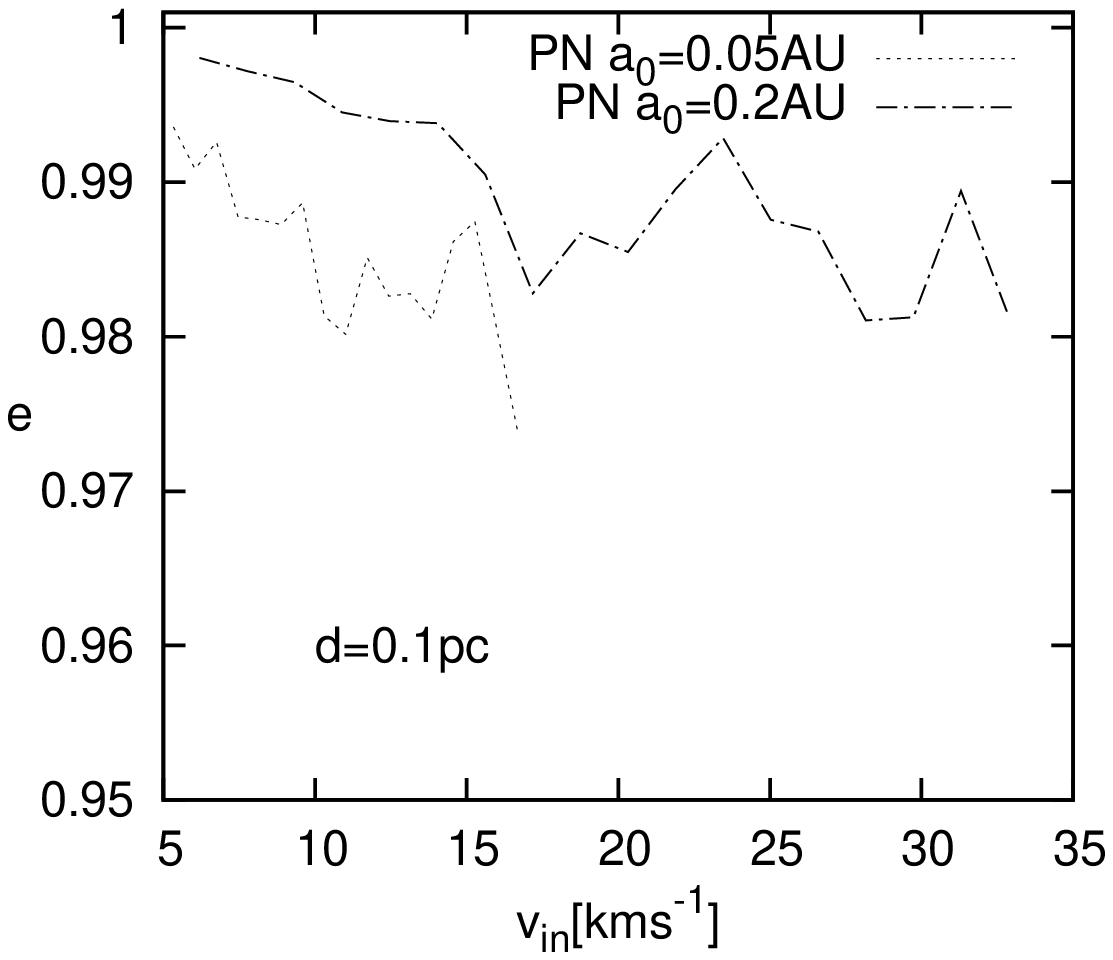}&
\includegraphics[angle=270,width=1.7in]{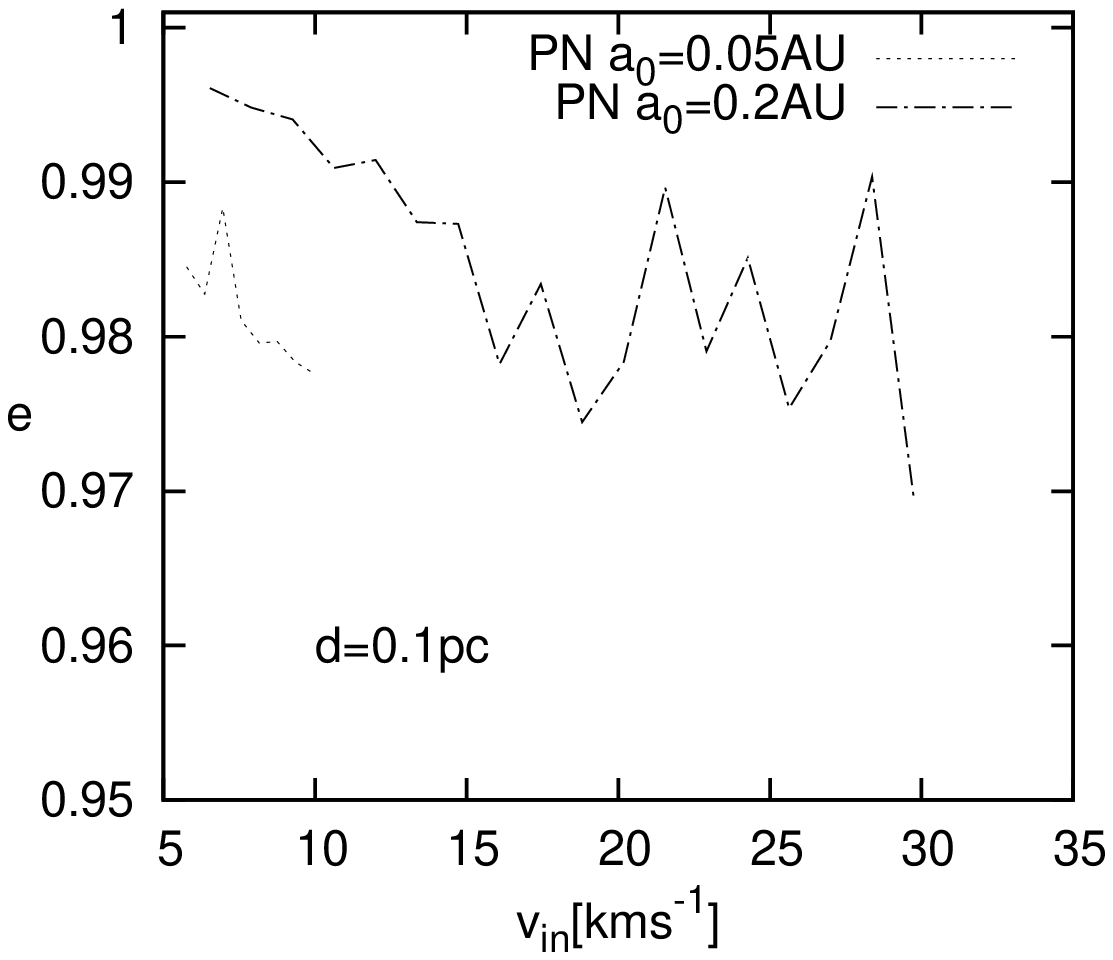}
\end{array}$
\caption{ eccentricity  
versus the initial velocity $v_{\rm in}$ of
 the initial orbit with respect to SMBH, for the orbits
  of bound stars with masses $3\msun$ (left panels) and
  $6\msun$ (right panels).
The eccentricity is average over initial binary orientation.
 Higher stellar masses and lower
  binary separations tend to reduce the orbital eccentricity of the
  captured stars. }
\label{fig5}
\end{center}
\end{figure}

\begin{figure*}
\begin{center}
\includegraphics[angle=270,width=7.3in]{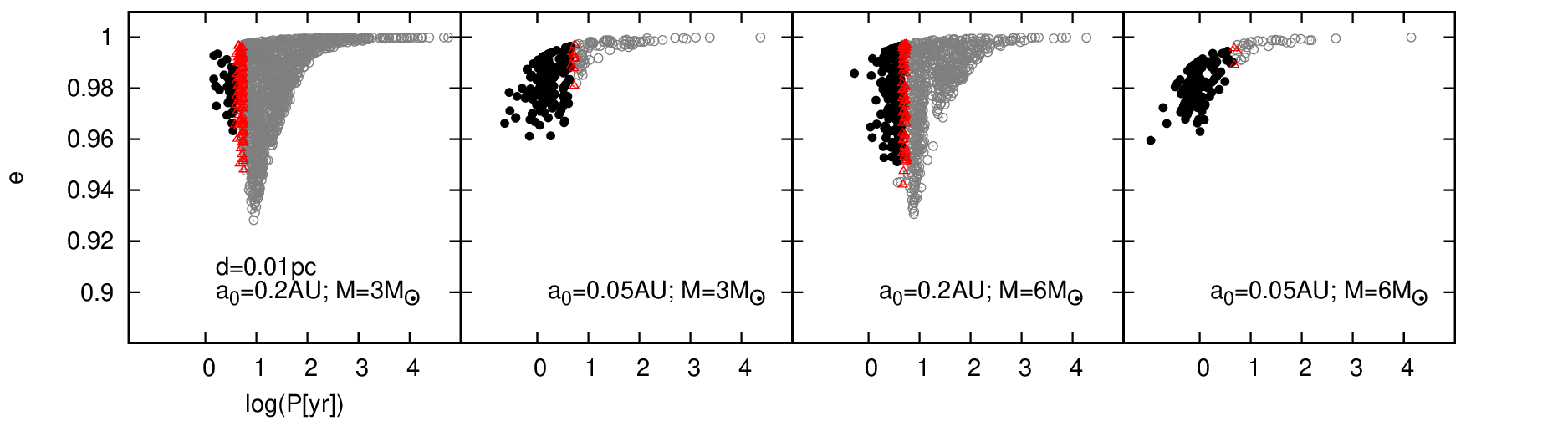} 
\includegraphics[angle=270,width=7.3in]{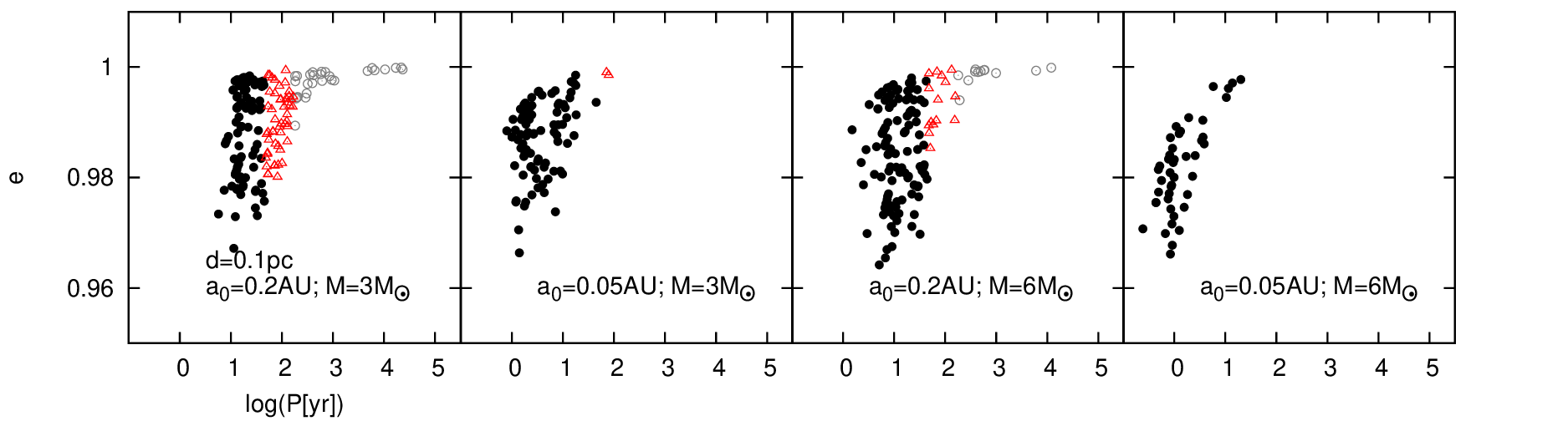} 
\caption{Eccentricity versus period for the orbits of the stars that 
remain bound to the SMBH at the end of the PN integrations. 
The grey open circles correspond to stars whose companion is also
  orbiting the SMBH. Stars whose companion has been ejected by the
  SMBH but remains bound to the Galaxy are represented by red triangles,
  while stars whose companion escapes as a HVS are indicated
  by filled circles.}
\label{fig6}
\end{center}
\end{figure*}

Figure~\ref{fig5} shows the dependence of the average final 
eccentricity of bound stars on $v_{\rm in}$. 
A decrease in $v_{\rm in}$ (and thus an increase in the eccentricity of 
the initial orbit) results in a larger final
eccentricity for the bound star, while larger  stellar masses and
initial binary separations decrease the eccentricity of the
captured star's orbit.  
Figure~\ref{fig6} displays the eccentricity instead as a
function of the orbital period of the stars that 
remain bound to the SMBH at the end of our PN simulations
(the Newtonian calculations look similar).  If we neglect the bound
stars produced by mergers, three different families can be identified:
\begin{itemize}
\item[1)] Bound stars whose companion also orbits the SMBH 
(grey open circles in Figure~\ref{fig6});
\item[2)] Bound stars whose companion has been ejected by the SMBH but 
remains bound to the Galaxy (red triangles in the figure);
\item[3)] Bound stars whose companion is a HVS   escaping the Galaxy 
  (filled circles in the figure). The approximate escape velocity from the Galaxy was chosen to be $1000 \kms$ 
  which  corresponds to stars  able to reach distances larger than $\sim 100 {\rm kpc}$ from the Galactic center  \citep{ke:08}.
\end{itemize}
Due to the dependence of the apoapsis distance on the ejection
velocity (see equation~\ref{ap}) the three families occupy distinct regions
on the $e-P$ plane. If more kinetic energy is deposited into the
ejected star, the apoapsis distance of the bound star is reduced.  In the case of equal mass binaries, the range of possible periods for
the three populations of bound stars are independent of the stellar mass 
and the binary separation.

On the other hand, in the case of unequal
mass binaries with  $q=M_{\rm b}/2M_1$ the apoapsis distance of
the bound star can be approximated by
\begin{equation} 
\label{ap2}
a_1 \sim \frac{GM_{\bullet}}{v_{\rm ej}^2}q,
\end{equation}
where $v_{\rm ej}$ is given by equation \ref{VEJ}  and for moderate mass ratios,  
the probability to be captured is almost independent by the
stellar mass.

Figure~\ref{fig6} shows that the stars with the highest values 
of $P$ and $e$  have  reached velocities  close to the escape 
velocity from the SMBH. 
Their companions also orbit the SMBH, but with much shorter
periods.  The region $e=0.92-0.95$, strongly populated for
$a_0=0.2\au$ and $d=0.01\rm pc$, is occupied by stars that were initially part of
binaries with the largest values  of $v_{\rm in}$.
For $d=0.1 {\rm pc}$  bound stars are mainly companions of ejected stars and the comparable region
is empty.

 Stars whose the companion is escaping the Galaxy have orbital  periods $\lesssim 50 {\rm yr}$. 
Many cases with orbital periods $\lesssim 10 \rm yr$ are also observed.

Finally, we note that in PN gravity the periapsis
distances of bound stars are about $0.1\au$ smaller than in the
Newtonian regime.
 \

\section{Collisions and Mergers}
In this section, we study the evolution of  binaries that are strongly 
perturbed during close passages to the SMBH, leading to stellar collisions 
and mergers. 
Stellar collisions due to either binary evolution or dynamical
interactions are thought to be the main mechanism for the production
of blue stragglers in star clusters \citep{CJ:84,BE:87,Leonard:89,MAT:90}. 
Similar processes may occur at the Galactic center,
producing a population of ``rejuvenated'' stars.
\citet{GE:03} proposed that the S-stars may be ``super-blue stragglers''  
formed by collisions of lower mass stars and/or tidal heating.
The apparent normality of their spectra \citep{EI:05} argues against  rejuvenation \citep{fig}, 
but given the uncertainties in the relaxation of merger products and the redistribution of angular momentum, 
it is difficult to draw firm conclusions as to their history.

It is possible that the S-stars are drawn from the young population observed in the stellar 
disk that extends inward to within $0.1~{\rm pc}$ of the SMBH, but while the S-stars are primarily 
B dwarfs, the young stars in the disk are mainly luminous Wolf-Rayet and OB supergiants and giants  \citep{PA:01}.
This limits the connection between the two stellar populations, as
O/W-R stars are typically more massive and shorter-lived than B types.
\citet{LBK} proposed that binaries scattered from the stellar disk on highly eccentric bound orbits around the SMBH
would be disrupted, ejecting one member as a HVS while leaving the other member bound. 
The inspiral of a cluster hosting an IMBH would generate stars on comparable orbits.

GL07 noted that the production of HVSs at the Galactic center could also
result in collisions: the SMBH
disrupts a binary, delivering an impulsive kick to one of the two stars
at some direction with respect to the orbital plane, and under
some circumstances the two stars will collide. We expect a dependence of
the collision probability on the initial binary orientation (unless
gravitational focusing dominates) and a correlation between the number
of HVSs produced and the number of collisions.  
However, GL07 concluded 
that the frequency of mergers is so low that no star in the Galactic
center is expected to have been produced by binary coalescence.

To investigate the rates of mergers and collisions using $N$-body techniques, 
we define the minimum impact parameter for a collision as $2R_*$,
and the two stars are assumed to merge when their relative speed upon
collision is lower than the escape velocity from their surfaces.

We find a much higher rate of coalescence than found by GL07,
with most close collisions occurring only after repeated encounters of the 
binary with the SMBH.
The mechanism that leads to the majority of the close encounters 
in our integrations is Kozai oscillations.

\begin{table*}
  \centering
  \caption{HVSs ($v_{\rm ej}>1000\kms$), Collision and Merger frequency(\%) for runs with $r_{\rm per} \le  6r_{\rm bt}$. 
    \label{tab2}} 
  \begin{tabular}{lllllllllllll}
    \hline 
    \hline
    & \phantom & \multicolumn{3}{c}{\phantom} & \multicolumn{4}{c}{post-Newtonian (Newtonian); $d=0.01$pc} & \multicolumn{4}{c}{\phantom}  \\
    \hline 
    & \phantom & \multicolumn{3}{c}{1 orbital period} & \multicolumn{4}{c}{5 orbital periods} & \multicolumn{4}{c}{60 orbital periods}   \\ 
    \cline{3-5} \cline{7-9} \cline{11-13}\\ 
    $a_0[{\rm AU}]$, $M_{*}[{\rm M_\odot}]$& \phantom & HVSs   & Coll  & Merg & \phantom   & HVSs   & Coll    & Merg & \phantom  & HVSs  & Coll & Merg   \\ 
    \hline
    0.2, 3 &  \phantom  &  3.14 (3.00) & 2.54(2.18) & 1.50(2.15) & \phantom  & 3.65 (3.25) & 8.30 (9.10) & 5.49 (5.29)         & \phantom  & 4.31 (3.26) & 22.0 (20.5)          & 17.0(16.6)  \\ 
    0.2, 6 &  \phantom  &  15.2  (15.4) & 4.47(2.92) & 3.01(2.62) &\phantom   & 16.5 (16.7) & 18.3 (12.5) & 9.11 (9.59)         & \phantom  & 17.4 (16.8) & 30.6 (30.6)          & 27.5(27.4)  \\
    0.05, 3 & \phantom &  16.3  (16.8) & 8.05(9.74) & 5.97(5.45) & \phantom  & 21.7 (18.2) & 15.7 (23.4) & 14.4 (20.6)         & \phantom  & 22.4 (21.0) & 41.0(47.0)           & 40.1(43.8)  \\
    0.05, 6 & \phantom &  15.6  (16.2) & 13.8(16.2) & 13.8(13.3) &\phantom   & 15.9 (16.9) & 24.3 (30.2) & 23.9 (27.9)         & \phantom & 20.8 (20.2) & 47.5 (55.5)          & 47.1 (53.2) \\
    \hline
    & \phantom & \multicolumn{3}{c}{\phantom} & \multicolumn{4}{c}{post-Newtonian; $d=0.1$pc} & \multicolumn{4}{c}{\phantom}  \\
    \hline 
    & \phantom & \multicolumn{3}{c}{1 orbital period} & \multicolumn{4}{c}{5 orbital periods} & \multicolumn{4}{c}{60 orbital periods}   \\ 
    \cline{3-5} \cline{7-9} \cline{11-13}\\ 
    $a_0[{\rm AU}]$, $M_{*}[{\rm M_\odot}]$& \phantom & HVSs   & Coll  & Merg & \phantom   & HVSs   & Coll    & Merg & \phantom  & HVSs  & Coll & Merg   \\ 
    \hline
    0.2, 3 &  \phantom   & 29.0 & 1.66 & 1.00   & \phantom    & 30.0  & 5.00  & 3.60   & \phantom  & 30.0  & 18.6    & 17.1 \\ 
    0.2, 6 &  \phantom   & 31.6 & 4.33 & 3.33   & \phantom    & 35.6  & 12.0  & 9.33   & \phantom  & 37.0  & 20.3    & 16.3 \\
    0.05, 3 & \phantom  & 25.0 & 8.33 & 8.05   & \phantom    & 27.1  & 21.7  & 21.0   & \phantom  & 28.0  & 30.3    & 29.6 \\
    0.05, 6 & \phantom  & 13.3 & 18.3 & 16.0   & \phantom    & 14.3  & 27.0  & 24.3   & \phantom  & 14.7  & 40.0    & 37.3 \\
    \hline
  \end{tabular}
\end{table*}

\subsection{Kozai oscillations}

The perturbations on the inner binary's orbit caused by the gravitational interaction 
with the SMBH can result in periodic oscillations (Kozai cycles) of both the  internal binary eccentricity 
$e_{\rm b}$ and the mutual inclination $j$ \citep{KO:62}.
This occurs  when the initial inclination $j_{\rm in}$ is sufficiently large and when the stars approach the SMBH
at relatively large periapses ($\gap r_{\rm bt}$) .
More precisely, the perturbations from the SMBH on the inner binary must always be weak and $j_{\rm in}$ 
needs to satisfy the relation  $i_{\rm c} \le j_{\rm in} \le 180^\circ - i_{\rm c}$, where the critical angle $i_{\rm c}$ can be assumed
to be $\sim 40^\circ$ in the case of initially circular binaries and Newtonian gravity.  
Kozai cycles can result in a reduction of the periapsis separation, 
allowing the inner bodies to collide.
The period of the cycles can be written in terms of the masses of the three bodies, the eccentricity of the outer binary
$e$ and their semimajor axis as:
\begin{eqnarray}\label{kp}
\tau = \frac{P_{\rm out}^2}{P_{\rm{b}}} \left( 1-e^2\right)^{3/2} \left(\frac{M_{\rm b}+M_{\bullet}}{M_{\bullet}} \right) K (e_{\rm b},\omega_{\rm b}, j_{\rm in}) \nonumber \\
 \simeq \frac{P_{\rm out}^2}{P_{\rm{b}}} \left( 1-e^2\right)^{3/2}~~~~~~~~~~~~~~~~~~~~~~~~~~~~~~~~~~~
\end{eqnarray}
where $P_{\rm b}$ is the periods of the inner  binary , $P_{\rm out}$ the external period, $e_{\rm b}$ and $\omega_{\rm b}$ 
the inner binary eccentricity and argument of periapsis, and $K$ is generally of order unity \citep{FORD}.
Writing 
\begin{eqnarray}
P_{\rm b} = 2 \pi \left(\frac{a_0^3}{G~M_{\rm b}} \right)^{1/2}~~~~~~~~~~~~~~~~~~~~~~~~~~~~~~~~~\nonumber \\
=4.5\times10^{-3} \left[  \left( \frac{a_0}{0.05 \rm{AU}}\right)^3 \left( \frac{6M_{\odot}}{M_{\rm b}} \right)   \right]^{1/2} \rm{yr} \nonumber
\end{eqnarray}
and
\begin{eqnarray}
P_{\rm out}=  2 \pi \left(\frac{[d/(1+e) ]^3}{G~M_{\bullet}} \right)^{1/2}~~~~~~~~~~~~~~~~~~~~~\nonumber \\
=\frac{44.4}{(1+e)^{3/2}} \left[  \left( \frac{d}{0.01\rm{pc}}\right)^3 \left( \frac{4\times 10^6 M_{\odot}}{M_{\bullet}} \right)   \right]^{1/2} \rm{yr} \nonumber
\end{eqnarray}
the Kozai period becomes
\begin{eqnarray}
\tau \simeq  \frac{2 \pi}{\sqrt{G}} \frac{M_{\rm b}^{1/2}}{M_{\bullet}} \left( \frac{d}{a_0^{1/2}} \right)^3  \frac{\left( 1-e^2\right)^{3/2}}{\left(1+e\right)^3} ~~~~~~~~~~~\nonumber \\
=4.4 \times 10^5 \left( \frac{a_0}{0.05\rm{AU}} \right)^{-3/2} \left( \frac{M_{\rm b}}{6\rm{M_{\odot}}} \right)^{1/2}  \nonumber \\
\times \left( \frac{M_{\bullet}}{4\times 10^6\rm{M_{\odot}}} \right)^{-1}\left( \frac{d}{0.01\rm{pc}} \right)^{3}  \frac{\left( 1-e^2\right)^{3/2}}{\left(1+e\right)^3} \rm{yr}. \nonumber
\end{eqnarray}
Thus, in our integrations, the 
Kozai period is between  $\sim 2$ and $ \sim 120$ times longer than
the orbital period around the SMBH, and the full effect of the Kozai cycles
is only felt after several revolutions.

If the orbit is initially circular, the maximum eccentricity achieved during a Kozai cycle is 
\begin{equation}\label{emax}
e_{\rm b,max}=\left[1-(5/3)\cos^2 j_{\rm in}\right]^{1/2}~,
\end{equation}
the amplitude of the oscillations depending only on the initial inclination of the inner orbit 
relative to the external perturber. 
The eccentricity of the outer binary remains roughly constant
throughout the evolution, because
its variation is caused by octupole interactions, which are weaker than the quadrupole interactions
driving the oscillations of the inner binary's  eccentricity \citep{TAK}.

Kozai resonances can be  suppressed by additional sources of apsidal precession
such as tides and relativistic  precession in the inner binary. 
The inclusion of PN effects between the two stars  damps  the oscillations
of the inner orbital eccentricity, eventually  causing a decrease in the number of observed collisions and
mergers \citep{HO:97,BL:02}.

\subsection{Results}

The results of our runs show, for what we believe to be the first time, 
that binary break-up is not a prerequisite for physical collisions,
and that the collision probability increases strongly with time: 
increasing the simulation time (and thus the number of revolutions around the SMBH) 
results in a dramatic increase in the number of collisions.  
The first point is of
particular relevance for binaries with high values of $r_{\rm per}$, for
which the SMBH more gradually reduces the internal binary angular
momentum. This would allow the two stars to closely interact through mass
transfer and eventually merge while they are still bound to each
other. 
We note that the first encounter with separation $\leq 2R_*$
usually occurs just after the closest approach of the binary to the
SMBH, when the orbital parameters suddenly change (see
Figure \ref{fig2}).  
With regard to the second point: allowing the orbit
to make only $2-3$ extra revolutions around the SMBH increases the
collision probability by up to three times with respect to shorter
integrations (GL07).  
This is a consequence of the fact 
that the Kozai period is longer than the period of revolution around the SMBH.

\begin{figure*}
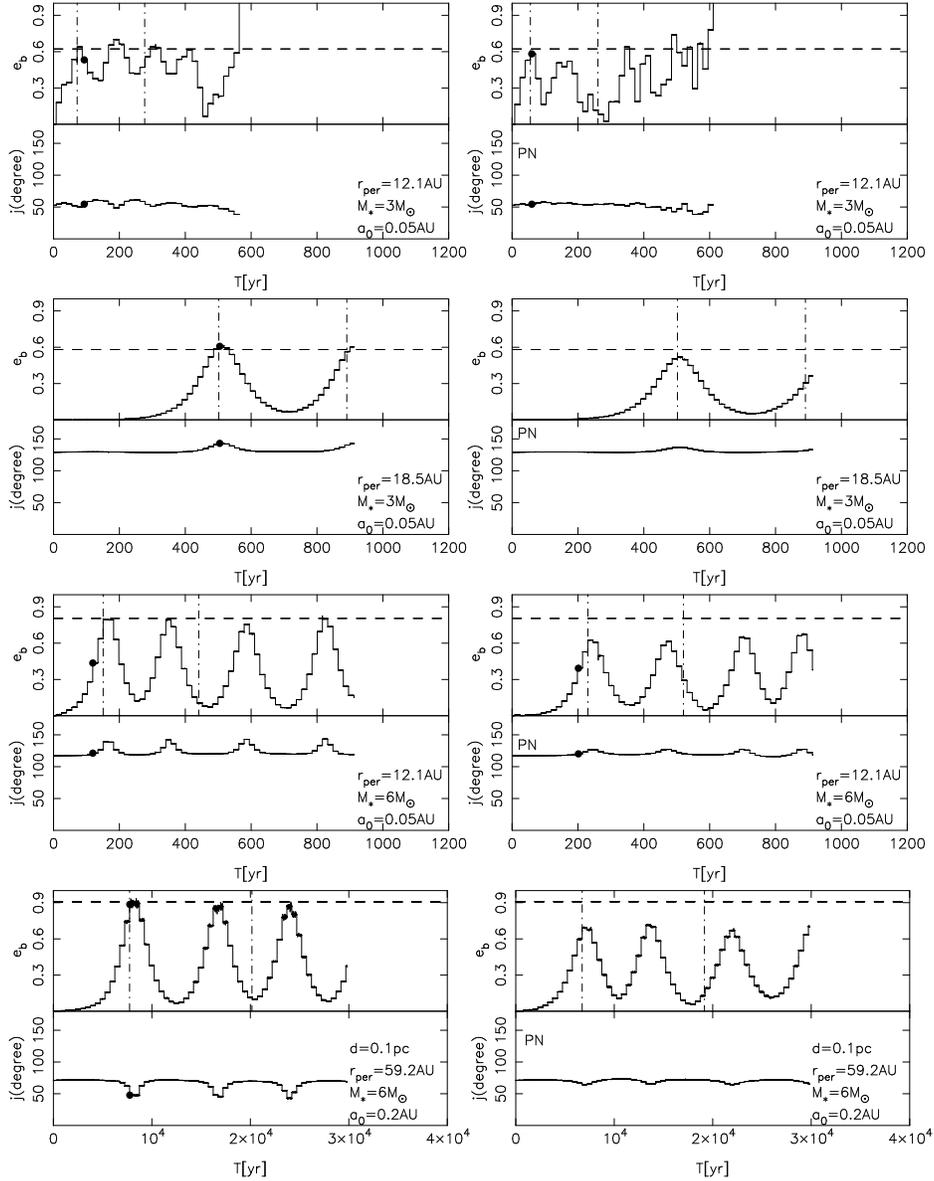

\begin{center}
$\begin{array}{ll}
\includegraphics[angle=0,width=2.4in]{fig7a.ps}
\includegraphics[angle=0,width=2.4in]{fig7b.ps} \\
\includegraphics[angle=0,width=2.4in]{fig7c.ps}
\includegraphics[angle=0,width=2.4in]{fig7d.ps} \\
\includegraphics[angle=0,width=2.4in]{fig7e.ps}
\includegraphics[angle=0,width=2.4in]{fig7f.ps} \\
\includegraphics[angle=0,width=2.42in]{fig7g.ps}
\includegraphics[angle=0,width=2.42in]{fig7h.ps} 
 \end{array}$
\caption{Evolution of the internal binary eccentricity and the mutual inclination $j$
for different  simulations in Newtonian (left panels) and PN (right panels) runs.  
In the cases shown, stars experience multiple collisions, and the 
time  at which they finally merge is marked by a filled circle.
For each periapsis distance considered , the corresponding 
Newtonian and PN integrations were started from the same initial conditions.
The top three rows have $d=0.01$pc and the bottom panels have $d=0.1$ pc.
The horizontal dashed lines 
marks the \emph{classical} value $e_{\rm b,max}$ given in equation (\ref{emax}).
The vertical dot-dashed lines give the Kozai timescale.
When the periapsis is sufficiently large the 
binary undergoes Kozai resonance, periodically changing both $e_{\rm b}$ and $j$.
When PN terms are included  the amplitude of the oscillations
decreases,  resulting in a larger survival time for the binary
and a smaller probability for a collision. The bottom panels
show the evolution for systems with $d=0.1$pc, for which
the stars do not merge if PN corrections are included.} 
\label{fig7}
\end{center}
\end{figure*}

Figure~\ref{fig7} gives four representative examples of how the inner binary parameters change 
when repetitive collisions occur between the stars, with and without  PN corrections.  These cases are difficult to
treat with an $N$-body approach, but we may safely conclude that the stars
are likely to merge after a short time. 
When relativistic terms are included, the maximum value of the binary eccentricity reached in a Kozai cycle,
$e_{\rm b,max}$,  decreases. 
However, we found that the effect of relativistic precession
on the likelihood of collisions  and mergers 
observed in our simulations is weak.

In Table \ref{tab2} we list the fraction of observed collisions,
mergers, and HVS ejections after the first binary-SMBH encounter, after $5$ orbital
periods, and after $60$ orbital periods.  After the
first encounter, most of the binaries with small periapsis are
broken apart, depositing one star on a tight orbit around the SMBH
while ejecting the companion as a HVS. The frequency of HVS ejection
is nearly unchanged over time because the orbits with larger
periapsis do not get close enough to the SMBH to eject a member.
Such orbits, instead, suffer strong perturbations from the SMBH,
altering the internal orbital parameters of the binary. The larger the
periapsis, the larger the time required for the perturbations to
become important, and the collision and merger frequency increase
strongly with time. 
PN  terms have important consequences only for 
initially small binary separations since  the relativistic precession period is smallest
for these systems.  
However, as shown in Table \ref{tab2} 
the fraction of collisions is reduced  by no more than  $10\%$ for $d=0.01$pc,
 since for tight binaries even small eccentricities lead to close stellar encounters.

When the apoapsis is $d=0.1$pc the ratio between the timescale of relativistic precession in the binary 
and the   period of the Kozai cycles becomes much smaller suppressing the  Kozai resonance.
Writing the timescale of relativistic precession as:
\begin{eqnarray}
\dot{\omega}=\frac{3G^{3/2}M^{3/2}_{\rm b} }{a_0^{5/2}~c^2~(1-e^2_{\rm b})}~~~~~~~~~~~~~~~~~~~~~~~~~~\nonumber \\
=4.9\times 10^{-3} \left( \frac{M_{\rm b}}{6 M_{\odot}} \right)^{3/2} \left( \frac{a_0}{0.05 {\rm AU}} \right)^{-5/2} {\rm yr}^{-1} 
\end{eqnarray}
with c the speed of light,
the product $\dot{\omega}\tau$ of the initial configuration gives the relative strength of relativistic precession to that 
of the tidal field of the SMBH.
For $M=6 M_{\odot}$ and $r_{\rm per}\gtrsim 12 {\rm AU}$ , this product can be as large as $10^2$ 
and the Kozai cycles are completely suppressed \citep{FAB}.
The ejection probability, instead, increases  due to the larger eccentricity of the external orbit.
The combination of these effects causes a smaller fraction of collisions/mergers respect 
to the number of HVSs. 

\begin{figure*}
  \begin{center}
    $\begin{array}{cc}
    \includegraphics[angle=270,width=3.5in]{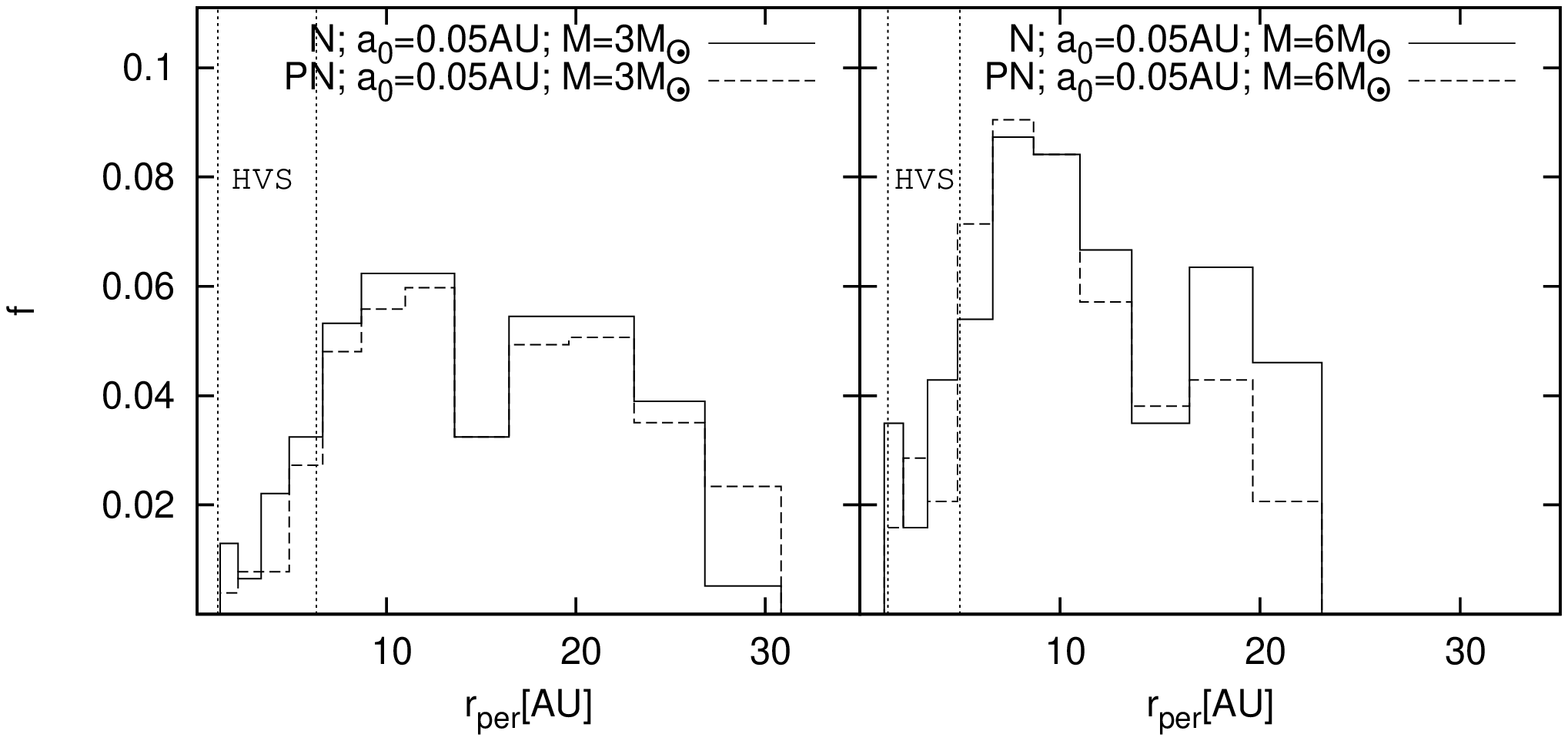} &
    \includegraphics[angle=270,width=3.5in]{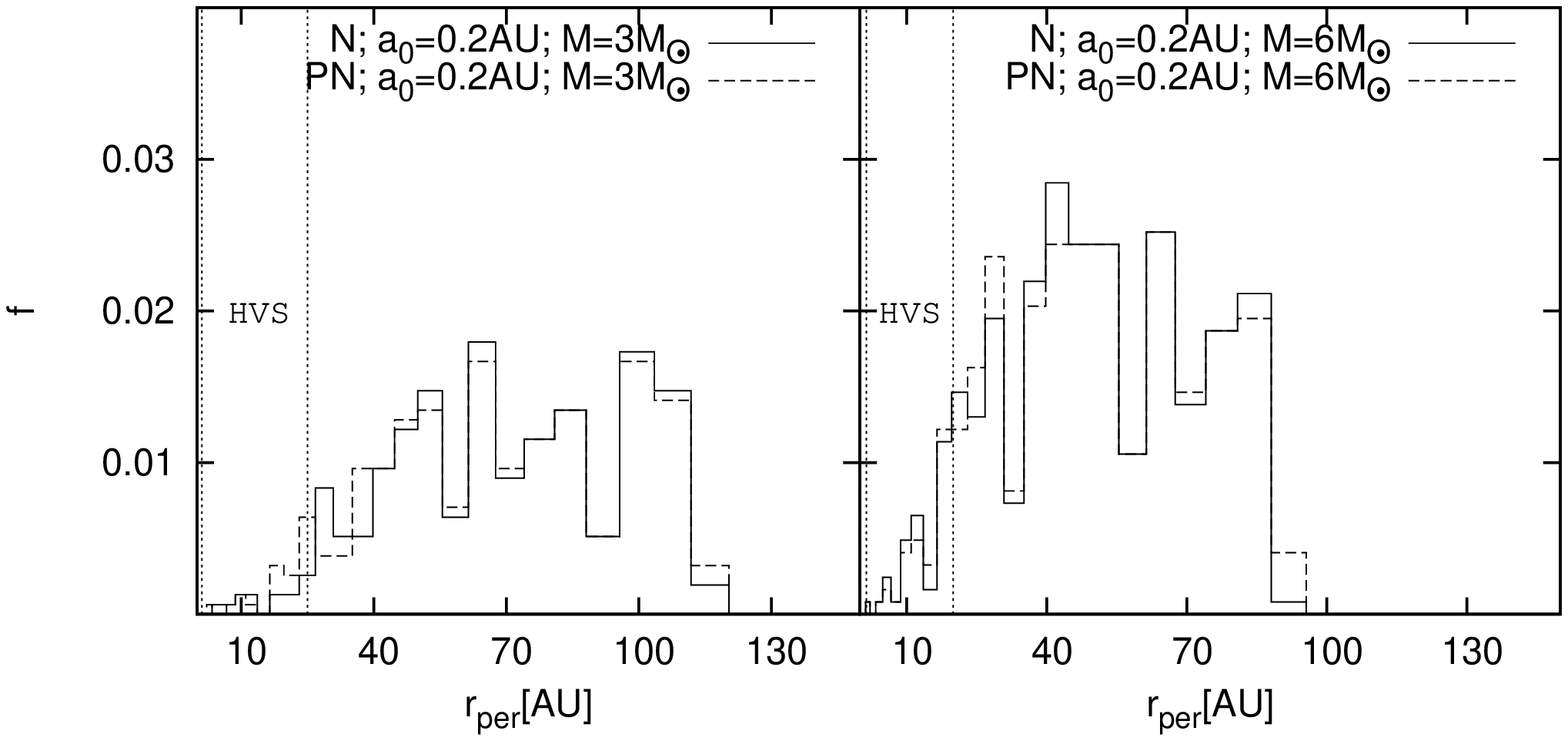} \\
    \includegraphics[angle=270,width=3.5in]{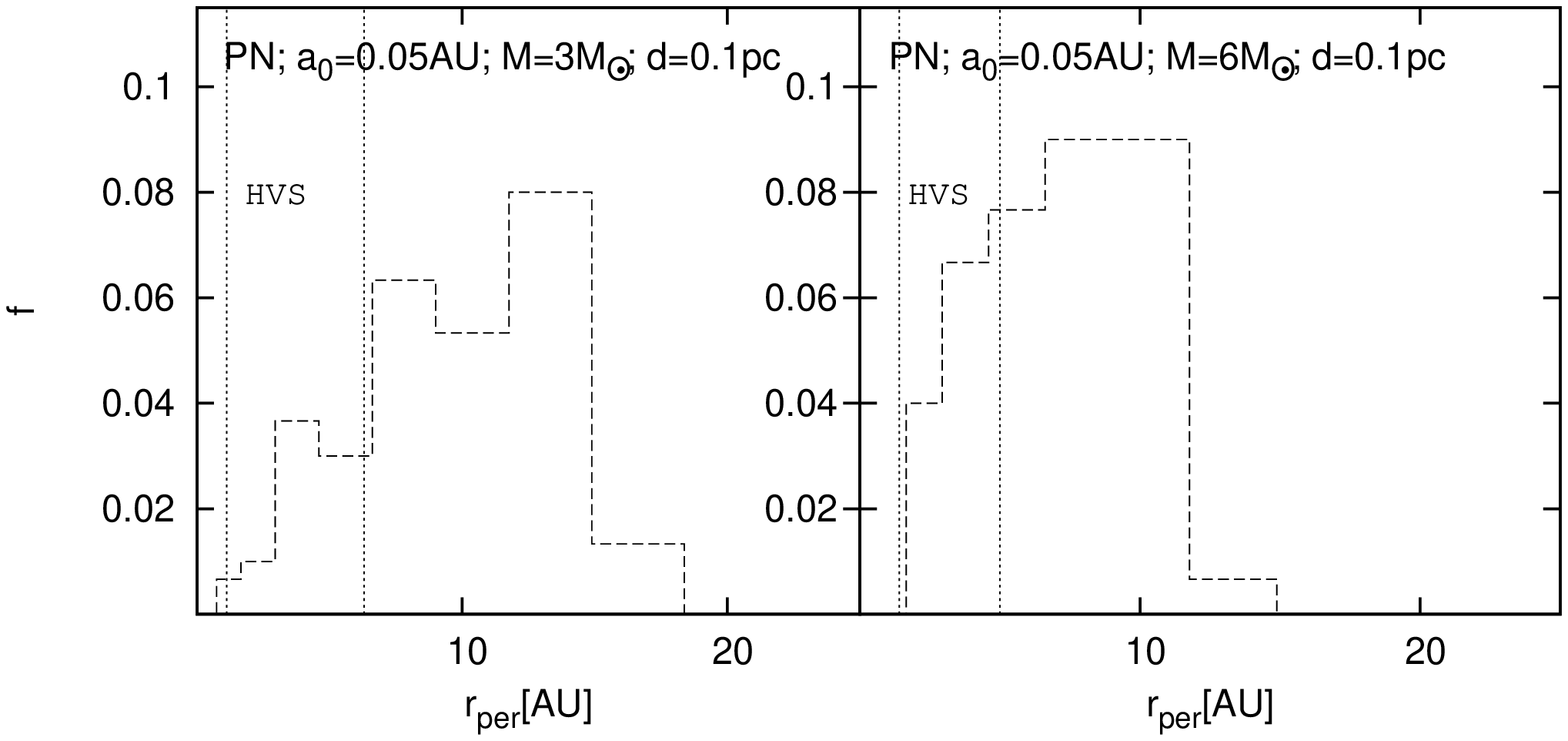}&
    \includegraphics[angle=270,width=3.5in]{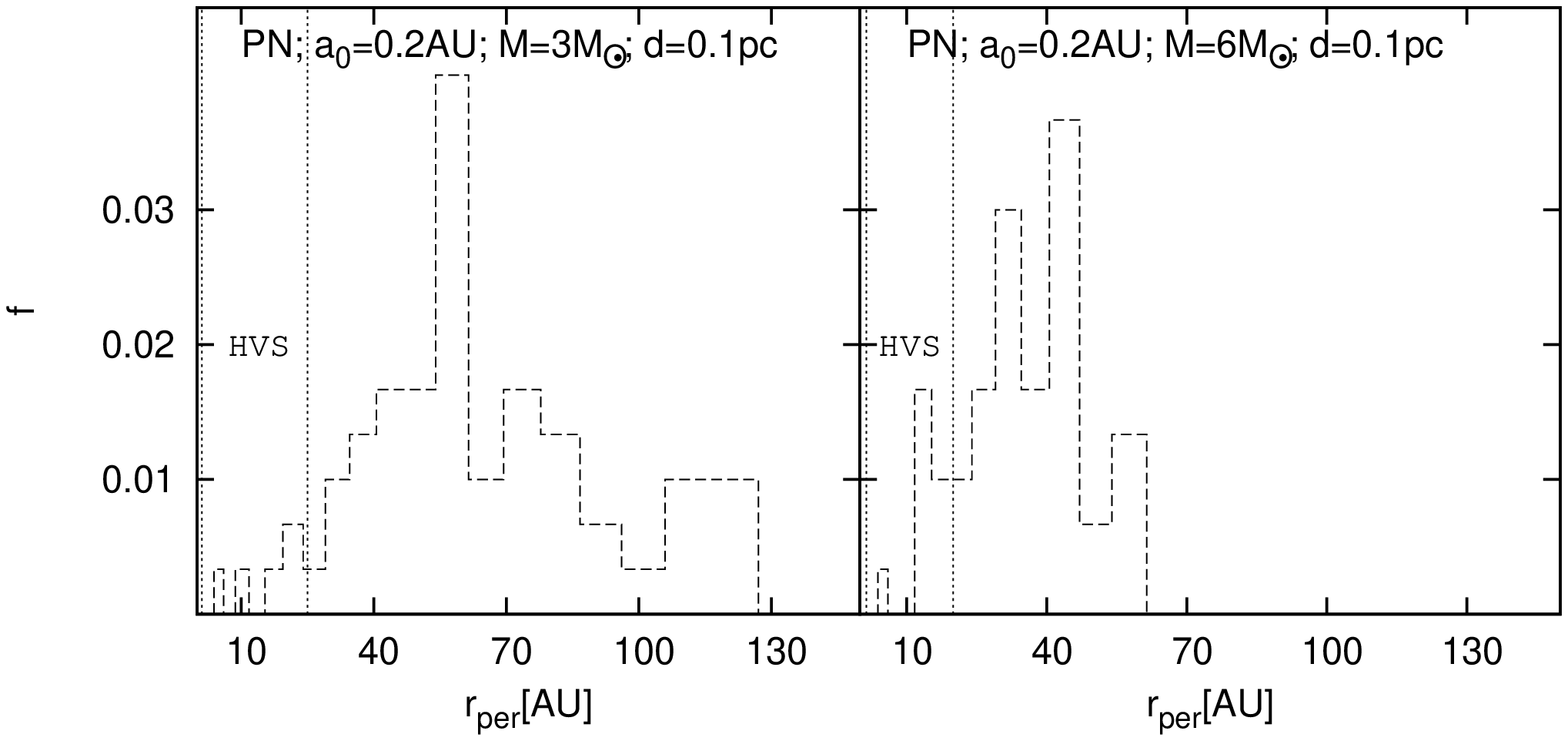}
    \end{array}$
    \caption{Fraction $f$ of orbits leading to stellar mergers as a function of the periapsis
      distance for PN and Newtonian simulations.
      The vertical dot-dashed lines bound the region where HVSs are
      produced.}    \label{IST}
  \end{center}
\end{figure*}

\begin{figure*}
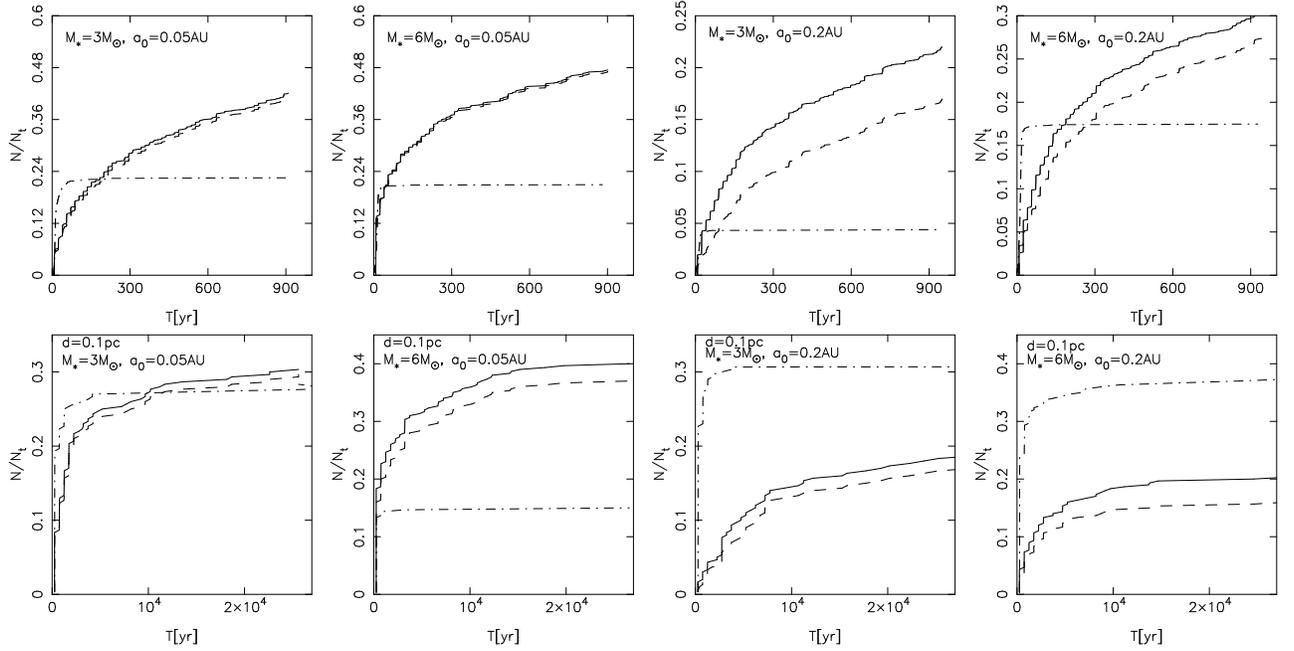

  \begin{center}
    $\begin{array}{cccc}
    \includegraphics[angle=0,width=1.6in]{fig9a.ps} &
    \includegraphics[angle=0,width=1.6in]{fig9b.ps} &
    \includegraphics[angle=0,width=1.6in]{fig9c.ps} &
    \includegraphics[angle=0,width=1.6in]{fig9d.ps} \\
    \includegraphics[angle=0,width=1.6in]{fig9e.ps} &
    \includegraphics[angle=0,width=1.6in]{fig9f.ps} &
    \includegraphics[angle=0,width=1.6in]{fig9g.ps} &
    \includegraphics[angle=0,width=1.6in]{fig9h.ps} 
    \end{array}$
    \caption{Cumulative fraction of stellar collisions (solid lines),
      mergers (dashed lines) and ejected HVSs (dot-dashed lines) in our PN integrations.
      Most of the HVSs are ejected during the first encounter with the SMBH;
      collisions and mergers, on the other hand, occur more frequently later after the inner binary's parameters 
      have evolved due to the  gravitational perturbations induced by the SMBH.}
    \label{cum}
  \end{center}
\end{figure*}

Figure \ref{IST} displays the fraction of stellar mergers as a function of the 
periapsis distance $r_{\rm per}$, with most cases occurring for
$r_{\rm per} > r_{\rm bt}$ where the 
Kozai mechanism becomes active.
Figure \ref{cum} shows the cumulative fraction of stellar collisions,
mergers and ejected HVSs in our PN integrations.
Most of the HVSs are ejected during the first periapsis passage.
Collisions and mergers, instead, typically occur later due to Kozai resonance, 
on a typical time scale $\sim \tau/2 $, where $\tau$ is given by equation~(\ref{kp}).

We note that strong mass transfer and/or mergers can lead
to the formation of a rejuvenated star
\citep{VAN:98,DT:07}.

The implications of these results for the actual rate of stellar mergers  
at the center of the Galaxy will depend on the dominant mechanism that drives  
binaries toward the SMBH.
If binaries are scattered inward by ``massive perturbers''
\citep{PHA}, most of their orbits will be weakly bound or unbound  with  
respect to the SMBH, with apoapses of order  parsecs, and will only
encounter the SMBH once before being scattered onto different orbits.
Given the low probability which we find for collisions during the first encounter with the SMBH for $d=0.1$pc, 
we do not expect a significant merger rate unless the binary population at the
Galactic center is biased  toward low values of  $a_0$.
Moreover,  if the binary is initially unbound, the merger product is likely to 
be ejected at low velocities and it will neither remain bound 
(unless substantial mass loss occurs) nor become a HVS.  
Alternatively, mergers in this scenario may result from disruption of triples 
that can leave binaries on bound orbits with small apoapse
\citep{PER:09a}.  
Obtaining an explicit value for the merger rate would be difficult in this case
given the large uncertainties about the population of multiple stellar systems in the Galactic center.

Alternatively, the binaries may form closer to the SMBH, perhaps in
a disk, and migrate inward.
In this case, the rate of collisions/mergers, as given in Table 3, 
would be comparable to the ejection rate of HVSs, i.e. 
$\sim10^{-5} {\rm yr^{-1}}$.
However, given the  strong  dependence of the collision probability on
the distance of closest approach to the SMBH,  the merger rate
 will be strongly correlated with the periapse distribution 
of the infalling binaries.
The generation of massive new stars through binary-SMBH encounters 
in this scenario could possibly explain
the presence of the stars that appear to be the youngest among the 
S-star group at the Galactic center  (see also \citet{PER:09a} and  
\citet{PF:09} for discussion on rejuvanting stars through Kozai-oscillation  
induced mergers).

In our simulations, the distribution of the semimajor axes and eccentricities of the
captured stars are fixed by the initial distance $d$ of the binary we
adopt, and therefore is completely arbitrary. 
Nevertheless, as previously discussed, the apoapses found in our runs are consistent with those
predicted assuming the binary progenitors of HVSs were scattered on eccentric orbits  
bound to the SMBH.
The orbital eccentricities we find ($e > 0.95$), it should be noted,  are larger than
those observed for the S-stars at the Galactic center, which are
consistent with a thermal distribution (i.e., $N(e) \propto e$).
\citet{PA:09} have shown how perturbations from a dense cluster of
$10\msun$ stellar black holes, expected to accumulate around the SMBH
due to mass segregation \citep{HA:06}, tend to randomize orbits with
high eccentricity in $\sim ~\rm{20Myr}$. Alternatively, if the
binaries are transported to the center by the inspiral of a massive
cluster containing an IMBH, interactions with the same IMBH will
thermalize the eccentricities in $\sim ~\rm{1Myr}$ \citep{MA:09}.
These processes of orbital randomization
result also in the evolution of the periapsis distance
$r_{\rm per}$, but on a timescale much larger than that required for the binary systems to collide.
We note, finally, that the presence of external forces on the 
inner binary  as well as tidal friction between the stars
could suppress the Kozai resonances reducing the  number of
stellar collisions.

\subsection{Effect of the Initial Binary Orientation}
Based on the results of our simulations, we find no strong dependence of the collision and/or merger probability during the first SMBH passage on the 
initial binary orientation.  To quantify our results, we use the Rayleigh (dipole)
statistic $\zeta_{\rm c}$ ($\zeta_{\rm m}$) 
defined as the length of the resultant
of the unit vectors $l_i,~i=1,...,N,$ where $l_i$ is perpendicular to
the initial internal orbital plane of the $i^{th}$ binary system whose
component stars collide (merge) during the integration and $N$ is
the number of collisions or mergers.
Here the subscripts $c$ and $m$ refer to 'collision' and 'merger' respectively.

  For an isotropic distribution
$\zeta \sim \sqrt{N}$, while $\zeta \sim N$ if some preferential
direction exists \citep{RA:19}.

In order to test the agreement with an isotropic distribution we used
a Monte Carlo approach: for each set of simulation parameters we generated a set of
$1000$ samples, each containing $N$ randomly oriented orbits.  Then,
for the whole sample we computed the average of $\zeta$ and the
corresponding standard deviation and we compared the resulting values
with those obtained for the colliding/merging binaries.

\begin{table*}{\small}
\centering
\caption{Raleigh statistics
  $\zeta_c$ and $\zeta_m$ and values expected for an isotropic
  distribution for collisions ($EV_c$) and mergers($EV_m$) after the
  first binary-SMBH encounter. } 
\begin{tabular}{llllll}
\hline 
\hline
&  \multicolumn{3}{c}{Post-Newtonian(Newtonian); d=0.01pc} \\ 
 \cline{2-5}
  \\ $a_0[AU]$, $M_{*}[M_{\odot}]$ & $\zeta_c/N$ & $EV_c$ & $\zeta_m/N$ &{$EV_m$} \\
  \hline
0.2, 3  & 0.33 (0.30) & 0.18$\pm$0.15 &  0.25 (0.29)& 0.23 $\pm$ 0.18  \\ 
0.2, 6  & 0.19 (0.23) & 0.17$\pm$0.14 &  0.21 (0.25)& 0.20 $\pm$ 0.16  \\
0.05, 3 & 0.20 (0.18) & 0.12$\pm$0.10 &  0.26 (0.21)& 0.13 $\pm$ 0.11  \\
0.05, 6 & 0.16 (0.14) & 0.10$\pm$0.08 &  0.21 (0.15)& 0.11 $\pm$ 0.09 \\
\hline
&   \multicolumn{3}{c}{Post-Newtonian; d=0.1pc}  \\ 
 \cline{2-5}
  \\ $a_0$, $M_{*}$ & $\zeta_c/N$ & $EV_c$ & $\zeta_m/N$ &{$EV_m$} \\
  \hline
0.2, 3  & 0.45  & 0.38$\pm$0.26 &  0.46   & 0.43 $\pm$ 0.27  \\ 
0.2, 6  & 0.14  & 0.17$\pm$0.14 &  0.30   & 0.30 $\pm$ 0.24  \\
0.05, 3   & 0.28  & 0.20$\pm$0.16 &  0.33   & 0.20 $\pm$ 0.17  \\
0.05, 6 & 0.14  & 0.14$\pm$0.11 &  0.16   & 0.15 $\pm$ 0.12 \\
\hline\\ \\
\label{rayl}
\end{tabular}
\end{table*}

Table \ref{rayl} gives the estimated values of $\zeta$ after the first
binary-SMBH encounter as well as the expected values for an isotropic
distribution. 
Assuming a confidence interval of $90\%$, most of the values shown
in Table \ref{rayl} are consistent with an isotropic distribution.

\begin{figure*}
\begin{center}
$\begin{array}{cc}
\includegraphics[angle=0,width=3.2in]{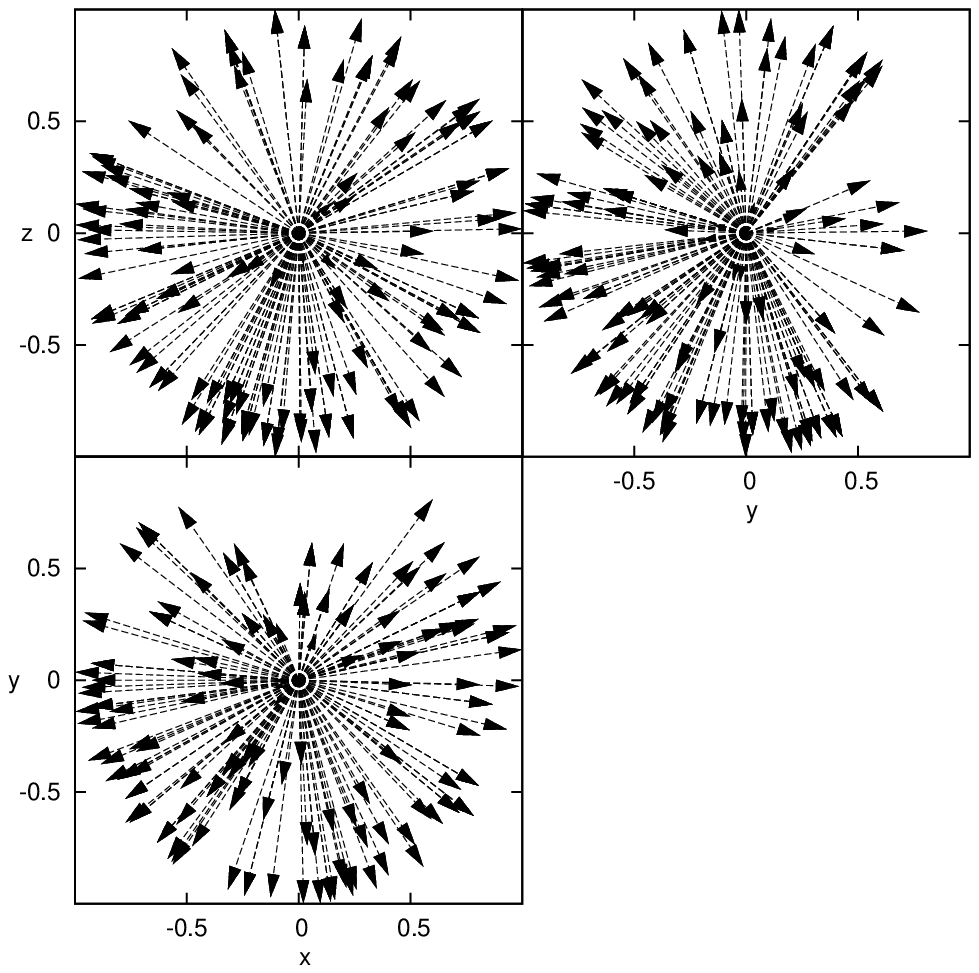} & \includegraphics[angle=0,width=3.2in]{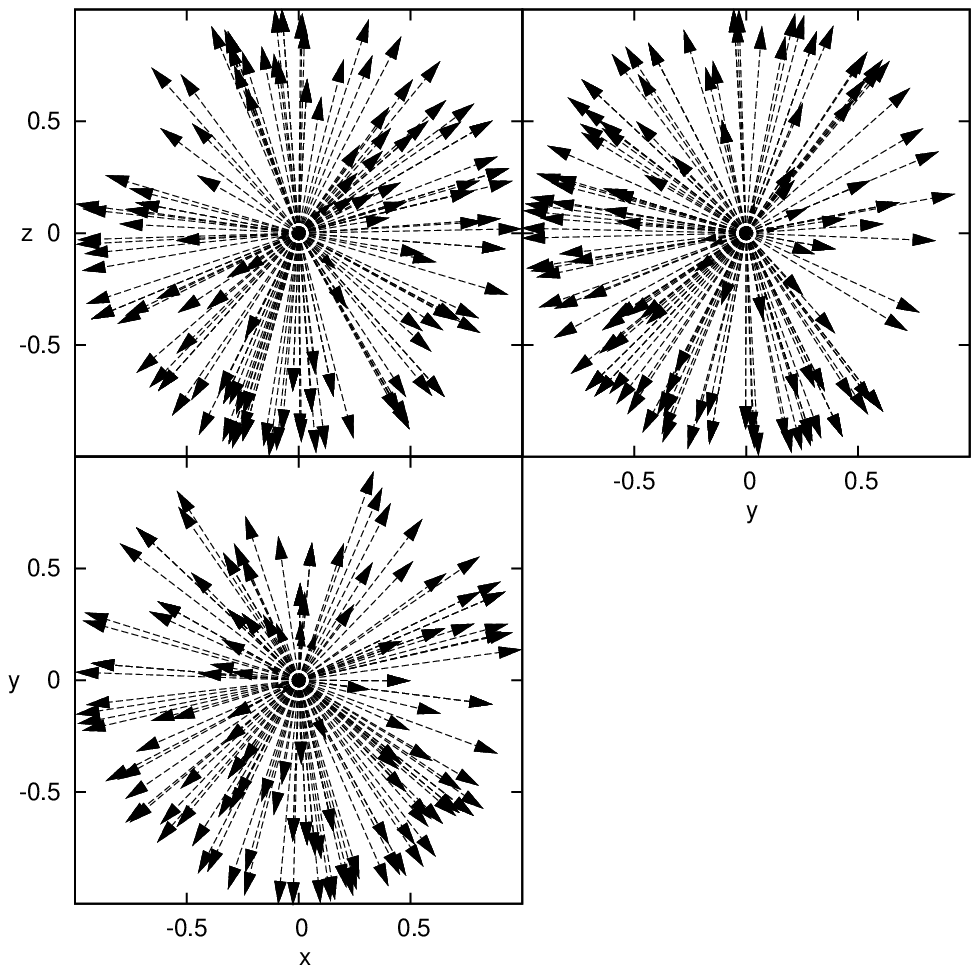}
 \end{array}$
\caption{Projection of the angular momentum eigenvectors for binaries with
  $a_0=0.05\au$ and $d=0.01$pc merging during the first encounter with the SMBH for
 PN (left panels) and Newtonian (right panels) runs. A slight dependence of
  the merger likelihood on the initial binary orientation is
  observed, with a deficit apparent for initial orbits whose
  angular momentum is aligned with the $+y$-axis.  The deficit is less significant in the Newtonian 
integrations.}
\label{angm}
\end{center}
\end{figure*}

We find values of $\zeta_m$ too large to be consistent with isotropy only for PN integrations
 where $a_0=0.05\au$ and $d=0.01$pc,
though for Newtonian integrations with the same apoapsis and $a_0=0.05\au$ ,
the values of $\zeta_m$ are quite large as well,  possibly indicating some degree of anisotropy.
In Figure~\ref{angm}, we plot the spatial projection of the
angular momentum eigenvectors
for the inner binary, showing that
when the angular momentum is aligned with the $+y$-axis, the
stars are unlikely to merge.  These binaries are retrograde and their
internal orbital plane coincides the orbital plane of the binary with
respect to the SMBH.  
In all the other cases we conclude that the
collision probability does not show any significant dependence on the
initial binary orientation during the first encounter with the SMBH.

\begin{figure}
\includegraphics[angle=0,width=3.2in]{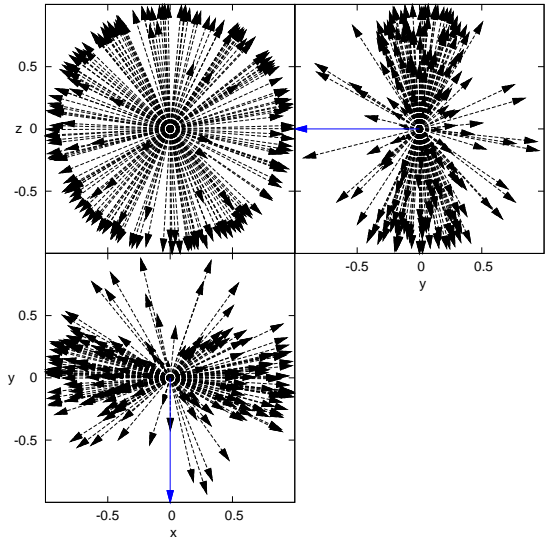} 
\caption{Projection of the angular momentum eigenvectors for binaries with
  $a_0=0.2\au$ and $d=0.01$pc merging between 10 and 60 orbital periods  for Newtonian  runs.  
  This shows how the Kozai mechanism selects highly inclined orbits (the external  orbital plane lies on the x-z plane and the
  direction of  angular momentum corresponding to the external orbit is given by a blue continue arrow).}\label{ang}
\end{figure}

At later times stellar collisions are mainly produced by the Kozai mechanism which is 
active only for large initial inclinations $j_{\rm in}$   .
This corresponds to a low  probability of collisions for binaries with the angular momentum  
aligned with either the $\pm y$-axis. 
For these cases our Rayleigh statistic results are still consistent with isotropy, but
this is found to be an artifact attributable to the dipole nature of the  
Rayleigh statistic.   
A more careful analysis  of the orbital distribution of binaries whose 
components merge at later times shows that, as expected, the population of 
colliding binaries is strongly biased toward large inclinations of the inner 
orbit with respect to the external orbital plane.
An illustrative example is given in fig \ref{ang} which 
shows the projection of the internal angular momentum vectors for binaries with
$a_0=0.2\au$ and $d=0.01$pc merging between 10 and 60 orbital periods.  
Even though the result of our statistical analysis in this case is consistent with isotropy,  
the  distribution of the angular momenta  is clearly   anisotropic with most of the  collisions occurring 
for large inclinations.
The discrepancy is   a consequence of  the dipole nature of the Rayleigh statistic:  
the majority of the  unit vectors $l_i$ have a quasi-antialigned counterpart giving a net contribution close to zero.

\section{Summary and Discussion}

We have carried out post-Newtonian numerical integrations of binary stars
on highly elliptical orbits around a supermassive black hole and
classified the final fates of both stars.
Our main conclusions can be summarized as follows:

1. If the binary orbit passes through the stellar tidal disruption 
radius $r_{\rm t}$ of the SMBH, both stars are likely to be disrupted 
during the initial passage, although there is a non-trivial chance that 
one star may remain outside the tidal disruption radius while the 
other passes within. 
 We found that after the binary internal  energy becomes positive
the stellar orbits remain close to the initial trajectory of the disrupted progenitor.
This demonstrates that HVSs can suffer strong tidal perturbations and mass loss before ejection.

2.  Binaries with periapses in the range $r_{\rm t}<r_{\rm per}< r_{\rm{bt}}$
with $\sim r_{\rm bt}$  the tidal breakup radius of the binary
are the main hypervelocity star progenitors.
In these cases the (hypervelocity) ejection probability is typically 
larger than $50 \%$, although the combination of wider separations 
($a_0=0.2 {\rm AU}$) and small apoapses ($d=0.01{\rm pc}$)  can
result in probabilities less than $30 \%$.

3. For tighter binaries (taken here to have separations of $0.05~{\rm AU}$), 
there is a strong inverse correlation between the periapse distance of 
the binary and the HVS ejection velocity, while
the dependence is present but weaker for wider binaries.

4. While many binaries perform several orbits around the SMBH before 
merging and/or colliding, the vast majority of HVSs were
 produced during the first passage by the SMBH.  
This is due to the strong dependence of HVS production 
on the periapse distance of the binary orbit: 
binaries that get sufficiently close to the SMBH are broken apart,
while those farther away will typically be unbound or collide.

5. The tidal break-up of equal-mass binaries by the SMBH results in a population of bound stars with 
 periods and eccentricities defined mainly by the fate of the companion star (Fig.~6).
Companions to escaping hypervelocity stars remain
bound to the SMBH with orbital periods $P\lesssim 50{\rm yr}$, 
but  also  extremely short  periods $P<{\rm 1 yr}$ are found.
A much larger range of periods ($10^2 \lesssim P \lesssim 10^5$) is instead 
possible for stars whose companion is also left orbiting the  SBH.

6. 
For sufficiently inclined orbits, the Kozai mechanism produces periodic 
oscillations of the inner binary eccentricity.  
This  reduces the internal binary periapsis allowing for close stellar encounters.
We find that the probability for collisions between the components
of the binary increases with time, resulting in substantially larger
numbers of mergers than in simulations that stop after the first passage.

7. When  the external orbital period is large compared to the period of 
relativistic precession in the inner binary the PN corrections  tend to 
suppress the oscillations reducing the probability of a collision which, however,
remains high in our integrations.
For initially bound orbits with $r_{\rm per}<6r_{\rm bt}$ the probability 
that the two stars collide is similar (for $d=0.1 {\rm pc}$ ) or much larger 
(for $d=0.01 {\rm pc}$) than that
of hypervelocity ejection.

\bigskip

Stars caught on highly eccentric orbits  around the SMBH   (with periapsis  $r_{\rm t}<r<2r_{\rm t}$)
can eventually spiral in as their orbital energy is converted into heat through tidal interactions during each periapse passage \citep[e.g.][]{AM:03}.
Such stars would have orbital properties generally consistent with the S-star population, but $N$-body 
calculations cannot accurately predict the properties of the merger products.  Instead, in a future work, we plan to
 study the evolution of binaries as they orbit the BH and collide or merge using a smoothed particle hydrodynamics
 (SPH) code, to determine the properties of the resulting stars, including their masses and angular momentum distributions.  
This should help to clarify whether they would be expected to possess normal spectra or features more commonly associated with blue stragglers.

Finally, our study provides concrete evidence that relativistic effects 
can have important consequences for the distribution of HVSs and the 
collision/merger probabilities of HVS progenitors.

\begin{acknowledgements}

We are extremely grateful to S. Mikkola who wrote the ARCHAIN algorithm
and who generously assisted us in using it.
We also thank T. Alexander, H. Perets, and H. Nakano for helpful 
conversations.  
J.A.F. was supported by NASA grants  HST-AR-11763.01 and 08-ATFP08-0093.  
D. M. and A. G. were supported by NASA grant NNX07AH15G
and by NSF grants AST-0807810 and AST-0821141.

\end{acknowledgements}

\end{document}